\documentclass[preprint,journal]{vgtc}                % final (journal style)
\ifpdf%                                % if we use pdflatex
  \pdfoutput=1\relax                   % create PDFs from pdfLaTeX
  \usepackage{graphicx}                % allow us to embed graphics files
  \DeclareGraphicsExtensions{.pdf,.png,.jpg,.jpeg} % for pdflatex we expect .pdf, .png, or .jpg files
\else%                                 % else we use pure latex
  \usepackage{graphicx}                % allow us to embed graphics files
  \DeclareGraphicsExtensions{.eps}     % for pure latex we expect eps files
\fi%

\graphicspath{{figures/}{pictures/}{images/}{./}} % where to search for the images

\usepackage{microtype}                 % use micro-typography (slightly more compact, better to read)
\PassOptionsToPackage{warn}{textcomp}  % to address font issues with \textrightarrow
\usepackage{textcomp}                  % use better special symbols
\usepackage{mathptmx}                  % use matching math font
\usepackage{times}                     % we use Times as the main font
\usepackage{cite}                      % needed to automatically sort the references
\usepackage{tabu}                      % only used for the table example
\usepackage{booktabs}                  % only used for the table example
\usepackage[T1]{fontenc}

\usepackage{color}
\usepackage{tikz}
\usetikzlibrary{shapes}
\usepackage{graphicx}
\usepackage[ruled,vlined]{algorithm2e}
\usepackage{hyperref}
\usepackage{float}
\usepackage{caption}
\usepackage{subcaption}
 \usepackage[moderate]{savetrees}

\usepackage{caption}
\usepackage{subcaption}
\usepackage{comment}

\usepackage[colorinlistoftodos]{todonotes} % comments in pdf

\ieeedoi{10.1109/TVCG.2019.2934402}

\newcommand{\REMOVE}[1]{\ignorespaces}
\newcommand{\ADD}[1]{#1}

\newcommand{\takeaway}[1]{
    \noindent\rule{\columnwidth}{0.5pt}
    \textbf{Take-away:} #1\vspace{-5pt}
    \rule{\columnwidth}{0.5pt}%
}

\newcommand{\datamodel}[1]{%
    \noindent\rule{\columnwidth}{0.5pt}
    \textbf{Model:} #1\vspace{-5pt}
    \rule{\columnwidth}{0.5pt}\vspace{5pt}%
}

\makeatletter
\if@todonotes@disabled

\else

\fi
\makeatother

\DeclareRobustCommand{\colorSquareBorder}[1]{{\tikz \draw [fill={#1}] rectangle (.7em,.7em);}}

\DeclareRobustCommand{\colorSquare}[1]{{\tikz \draw [fill={#1},draw=none] rectangle (.7em,.7em);}}

\definecolor{colorScaleMin}{HTML}{FCF7EC}
\definecolor{colorScaleMax}{HTML}{6B0000}% was 6B0000
\definecolor{colorScaleMCAL}{HTML}{bf812d}
\definecolor{colorScaleMCAR}{HTML}{35978f}
\definecolor{colorScaleACAL}{HTML}{8c510a}
\definecolor{colorScaleACAR}{HTML}{01665e}
\definecolor{colorScalePCAL}{HTML}{dfc27d}
\definecolor{colorScalePCAR}{HTML}{80cdc1}
\definecolor{colorScaleBA}{HTML}{d95f02}
\definecolor{colorScaleIC}{HTML}{7570b3}
\definecolor{colorScalePcomm}{HTML}{B22222}
\definecolor{colorScaleAcomm}{HTML}{000000}
\definecolor{colorScaleRed}{HTML}{FF0000}

\onlineid{0}

\vgtccategory{Research}
\vgtcpapertype{Design Study}

\title{CerebroVis: Designing an Abstract yet Spatially \\ Contextualized Cerebral Artery Network Visualization}

\author{Aditeya Pandey, Harsh Shukla, Geoffrey S. Young, Lei Qin, Amir A. Zamani, Liangge Hsu,\\ Raymond Huang, Cody Dunne, and Michelle A. Borkin}
\authorfooter{
\item Aditeya Pandey, Harsh Shukla, Cody Dunne,~and~Michelle~Borkin~are with Northeastern University. E-mail: \{pandey.ad,~shukla.h\}@husky.neu.edu, \{c.dunne, m.borkin\}@northeastern.edu
\item Geoffrey S. Young, Amir A.~Zamani,~Liangge~Hsu,~and~Raymond~Huang are with Brigham and Women's~Hospital.~E-mail:~\{gsyoung,~azamani, lhsu1, ryhuang\}@bwh.harvard.edu
\item Lei Qin is with the Dana-Farber Cancer Institute. E-mail: lqin2@partners.org
}

\shortauthortitle{Pandey \MakeLowercase{\textit{et al.}}: CerebroVis: Designing an Abstract yet Spatially Contextualized Cerebral Artery Network Visualization}

\abstract{Blood circulation in the human brain is supplied through a network of cerebral arteries.
If a clinician suspects a patient has a stroke or other cerebrovascular condition, they order imaging tests.
Neuroradiologists visually search the resulting scans for abnormalities.
Their visual search tasks correspond to the abstract network analysis tasks of browsing and path following.
To assist neuroradiologists in identifying cerebral artery abnormalities, we designed CerebroVis, a novel abstract---yet spatially contextualized---cerebral artery network visualization.
In this design study, we contribute a novel framing and definition of the cerebral artery system in terms of network theory and characterize neuroradiologist domain goals as abstract visualization and network analysis tasks.
Through an iterative, user-centered design process we developed an abstract network layout technique which incorporates cerebral artery spatial context.
The abstract visualization enables increased domain task performance over 3D geometry representations, while including spatial context helps preserve the user's mental map of the underlying geometry. 
We provide open source implementations of our network layout technique and prototype cerebral artery visualization tool.
We demonstrate the robustness of our technique by successfully laying out 61 open source brain scans.
We evaluate the effectiveness of our layout through a mixed methods study with three neuroradiologists.
In a \ADD{formative} controlled experiment our study participants used CerebroVis and a conventional 3D visualization to examine real cerebral artery imaging data to identify a simulated intracranial artery stenosis.
Participants were more accurate at identifying stenoses using CerebroVis (\REMOVE{odds ratio 2.5, }absolute risk difference 13\%).
\REMOVE{More broadly, we discuss the applications of our design approach to a general design paradigm we call \textit{Abstraction with Context}.}
A free copy of this paper, the evaluation stimuli and data, and source code are available at \href{https://osf.io/e5sxt/}{osf.io/e5sxt}.
}


\CCScatlist{ % not used in journal version
 \CCScat{K.6.1}{Management of Computing and Information Systems}%
{Project and People Management}{Life Cycle};
 \CCScat{K.7.m}{The Computing Profession}{Miscellaneous}{Ethics}
}

\teaser{
    \centering
    \includegraphics[width=\textwidth]{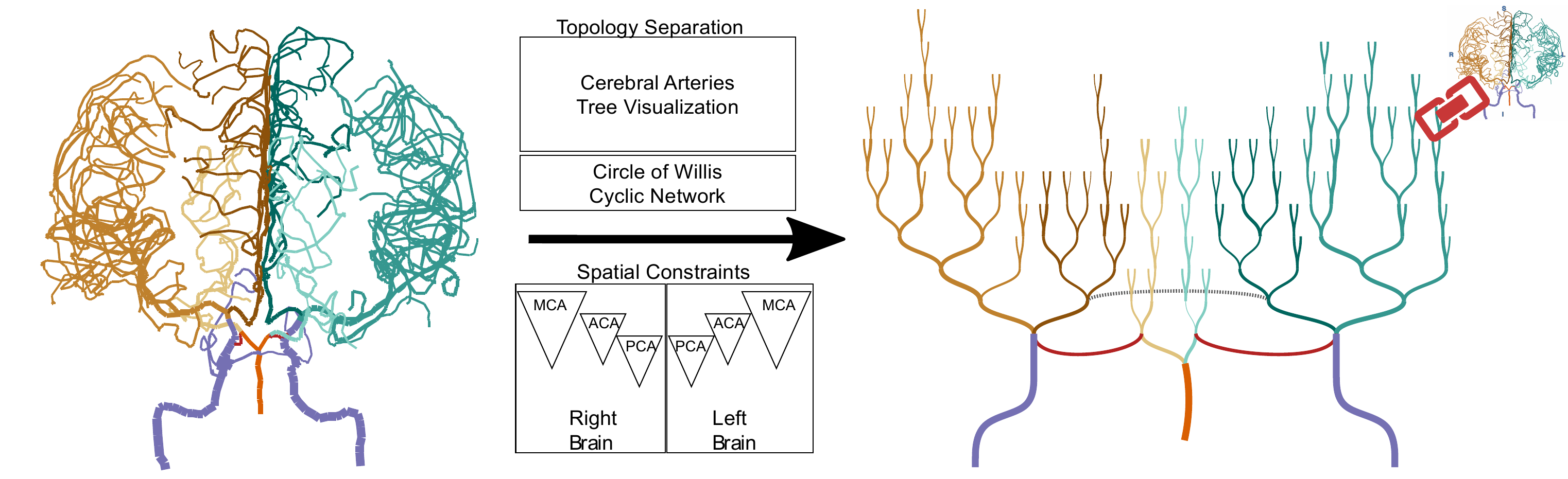}
    \vspace{-2.0em}
    \caption{
    CerebroVis is a novel network visualization for cerebral arteries.
    CerebroVis uses an abstract topology-preserving visual design which is put in spatial context by enforcing constraints on the network layout.
    Here we show the conversion of an almost symmetrical healthy human brain cerebral artery network from a 2D isosurface visualization (left) to CerebroVis (right).
    Each artery has the same categorical color in both views (see Sec.~\ref{brainvasc} for a legend).}
    \label{Teaser}
}

\begin{document}


\firstsection{Introduction} \label{sec:intro}

\maketitle

Arteries in the human brain form a network of blood flow, and a blockage or leakage in this network can lead to life-threatening cerebrovascular conditions such as a stroke or aneurysm.
Strokes alone are the fifth leading cause of death as well as a leading cause of serious long-term disability in the United States, and are globally the second leading cause of death after heart disease\cite{johnson2016stroke}.
Early detection and diagnosis of these conditions is essential for effective life-saving treatment.
Conventional diagnostics rely on an expert neuroradiologist identifying vascular abnormalities through examination of medical images (e.g., CTA, MRA).
This data is commonly rendered in 3D to assist the doctor with identification of the abnormalities.
However, prior research indicates that existing representations of the 3D cerebral arteries---e.g., isosurface, volume rendering, and Maximum Intensity Projection (MIPS)---introduce visual artifacts and task performance challenges such as overplotting/occlusion\cite{fishman2006volume}, false impression of geometry\cite{fishman2006volume}, and excessive artery bends.

In this design study, we present a novel 2D visualization of the cerebral artery system designed to assist doctors in the identification of cerebrovascular abnormalities.
Inspired by existing visualization research which has demonstrated the effectiveness of 2D representations for spatial search tasks in other medical imaging cases, e.g., cardiovascular arteries\cite{6065015} and connectomics\cite{neurolines}, we present a novel 2D abstract representation of the cerebral arteries.
To our knowledge, this is the first attempt to approach the cerebrovascular diagnostics tasks faced by neuroradiologists from the perspective of network science and using an abstract 2D visual encoding.

In this paper, we first offer a novel framing of cerebral arteries using network theory.
Next, we characterize the domain goals and present them as network analysis tasks.
In an iterative user-centered design with neuroradiology collaborators, we developed an effective abstract representation of the human cerebral artery network to assist neuroradiologists in identifying abnormalities.
We discovered that this new representation is most effective when spatial context is included to help users understand the novel representation.
In order to meet domain goals and satisfy design requirements, we developed a topology preserving network layout for cerebral arteries with spatial constraints to aid expert understanding.

We evaluate our new layout and the accompanying CerebroVis prototype in two ways: (1) assessing the robustness of the technique by examining 61 healthy brain scans\cite{wright2013digital} and (2) a mixed methods study with three neuroradiologists which included semi-structured interviews and a controlled experiment simulating intracranial stenosis diagnosis.
We found that our layout and implementation correctly visualizes all 61 brain scans, that neuroradiologists were more accurate at identifying stenosis with CerebroVis vs. a 3D visualization (\REMOVE{odds ratio 2.5,} absolute risk difference 13\%), and that neuroradiologists thought CerebroVis was easy to understand and a useful addition to their diagnosis toolbox.

\textbf{Contributions}: The primary contribution of this design study is a novel 2D, abstract, yet spatially contextualized cerebral artery network visualization and its open-source implementation in CerebroVis.
We also contribute a novel definition of the cerebral artery system in terms of network theory, as well as a network visualization task abstraction for cerebrovascular diagnostics.
Our evaluations with three neuroradiologists demonstrate the validity of our approach and confirm improved task performance for identifying cerebrovascular abnormalities as compared to a 3D visualization.
\REMOVE{Finally, we reflect on the generalizability of our \textit{Abstraction with Context} design paradigm for the benefit of the visualization community at large.}
\ADD{Finally, we reflect on the general aspects of our design paradigm for the benefit of the visualization community at large.}

\vspace{-0.3 em}

\section {Domain Background} \label{sec:domainbg}

Neuroradiology is a sub-specialty of radiology, the diagnosis of injuries and diseases with medical imaging, which focuses on conditions of the brain, spine, head, and neck.
In this work, we focus specifically on cerebrovascular diseases and related brain blood flow abnormalities.
In these diseases, which include stroke, aneurysm, and malformed vasculature, interrupted blood flow can deprive the brain of oxygen---with potentially lethal consequences.
In order to diagnose cerebrovascular diseases, a neuroradiologist visualizes imaging data to evaluate the arteries and identify physical abnormalities.
Abnormalities include vessel narrowing (stenosis), abnormal widening (fusiform aneurysm), berry shaped protrusion from normal arteries (thrombosis), and absence of normal vessels (occlusion).
Treatment for cerebrovascular diseases, sometimes with extreme time sensitivity, can include administering blood thinners or clotting agents to restore or stabilize blood flow and image-guided procedures to mechanically restore or alter artery structures.

A large number of specialized cerebrovascular imaging techniques exist\cite{10.1093/neuros/nyx325}.
The most common techniques are Compute Tomography Angiography (CTA)\cite{Aviv1975,Lev2736} and Magnetic Resonance Angiography (MRA)\cite{Aviv1975,Lev2736}.
In this paper we focus on MRA imaging due to its high resolution and open data set availability.
MRA uses magnetic radio-pulses for imaging with injected contrast dye to increase visibility and resolution of arteries.
The advantages of MRA as compared to CTA include higher quality imaging of soft tissues and no radiation (x-ray) exposure to the patient.
The disadvantages of MRA include the long acquisition time (typically close to an hour) and an unpleasant experience for patients with claustrophobia (patients must lie still in a narrow tube). 

\begin{figure}[tb]
    \includegraphics[width=\linewidth]{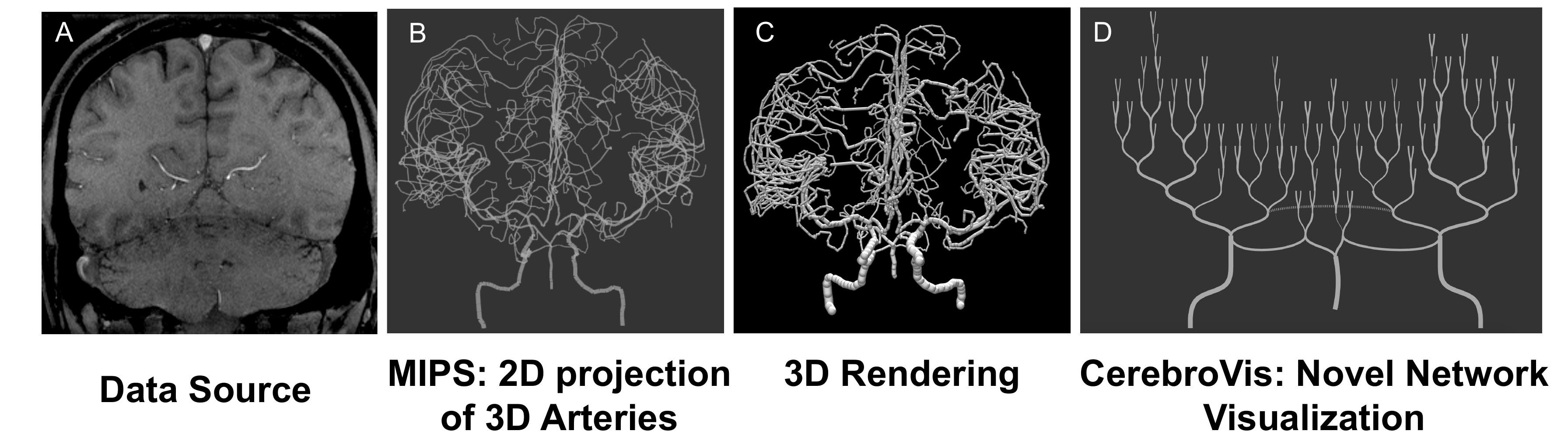}
    \caption{Existing cerebral artery visualizations: (A) raw image ``slices'', (B) MIPS, and (C) 3D  rendering.
    (D) Our CerebroVis visualization.} \label{visualizationtech}
\end{figure}

Fig.~\ref{visualizationtech} shows the three main methods used by neuroradiologists to visualize and analyze the cerebrovascular arteries.
First (A) we have a 2D image ``slice'' (orthogonal cut) through the raw imaging cube (typically in DICOM format).
Next (B) MIPS (maximum intensity projection)\cite{gacs1983ct,MIP} is a 2D projection of the 3D brain arteries.
However, the projection makes it difficult to delineate individual vessels \cite{fishman2006volume}. 
Third (C) is 3D rendering\cite{kubisch2012vessel,MIP} which preserves all spatial information and---through interaction---makes it possible to delineate individual arteries\cite{fishman2006volume}.
A key motivation for the development of our new method (D) was to clearly present each artery individually but obviate the need for interactivity.

\section{Network Model of the Cerebral Artery System}\label{brainvasc}

Here we present a novel definition of the cerebral artery system using network theory.
As will be discussed in Sec.~\ref{sec:CBVisDesign}, this network framing provided the essential insight and analytic strategies necessary for creating an effective 2D representation.
\vspace{-0.5 em}
\subsection{Networks and Trees Defined} 

A \textit{network} represents entities (\textit{nodes}) and the relationships between them (\textit{edges}).
In a \textit{directed} network, each edge has a source node and target and encodes a meaningful relationship direction. 
In an \textit{undirected} network, conversely, edges denote a bi-directional relationship.
A pair of reciprocal \textit{parallel} directed edges between two nodes can be modeled instead using a single undirected edge.
If a network contains both directed and undirected edges it is termed \textit{mixed}.
A \textit{path} is an alternating sequence of distinct nodes and edges which connect two nodes.
A network is \textit{connected} if there exists at least one path between every pair of nodes.
If a node is reachable from itself via a path, that path is termed a \textit{cycle} and the network is \textit{cyclic}.
A \textit{tree} is an undirected, connected, acyclic network. This means there exists only one path between any pair of nodes.
A \textit{directed tree} uses directed edges, but edge directionality is ignored when ensuring only one path exists between any pair of nodes. 
For hierarchical data a \textit{root} is assigned for the top level and edges in this \textit{rooted tree} model parent-child relationships.
\ADD{In a network, the \textit{degree} of a node is the total number of edges connected to the node.}
Nodes in the tree with a degree of at least two are called \textit{internal nodes}, while nodes with a degree of one are \textit{leaf nodes}. 
A tree in which each node has at most two children is called a \textit{binary tree}. 

\vspace{-0.2 em}

\subsection{Cerebral Arteries as a Network}

\begin{figure}[tb]
    \centering
    \includegraphics[width=.8\linewidth]{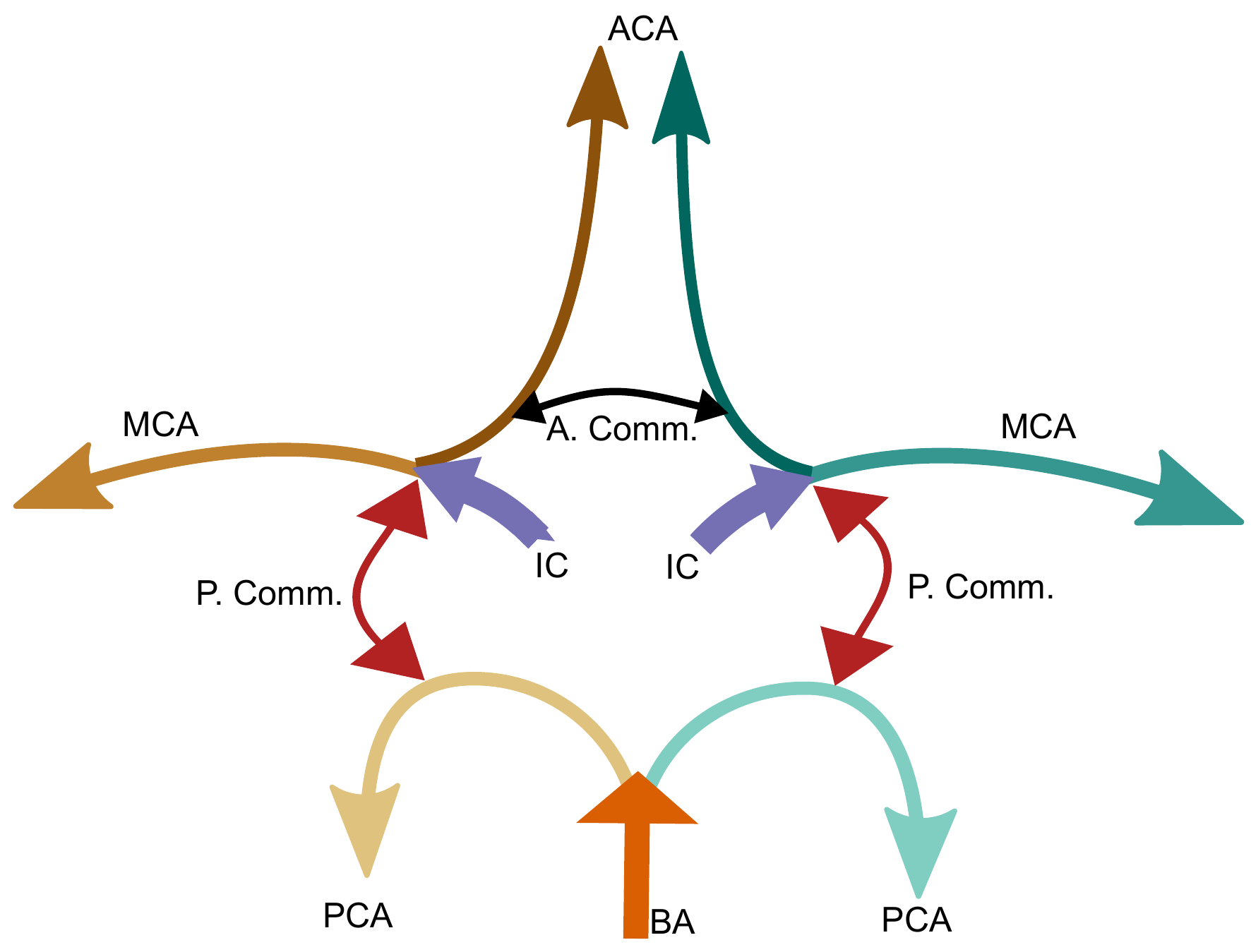}
    \caption{Diagram of the bi-directional Circle of Willis (CoW) cycle in the cerebral artery network.} \label{COW}
\end{figure}

The human circulatory system can be modeled as a blood flow network with edges representing vessels with variable amount of blood flow.
Arteries carry oxygenated blood away from the heart to the rest of the body including the brain.
Veins return deoxygenated blood back to the heart via the lungs.
In this paper we focus on arteries carrying blood to the brain.
Our network of interest has three main components: (1) arteries carrying blood from the heart to the brain, (2) the flow regulating Circle of Willis, and (3) arteries distributing blood inside the brain.
We discuss each of these below, including their diagnostic importance and a network theoretic data model.
The categorical colors used throughout the paper and components of interest are illustrated in Fig.~\ref{COW}.

\textbf{Arteries carrying blood to the brain:}
Four arteries supply blood to the brain: two \textit{internal carotid arteries (IC)} \colorSquare{colorScaleIC} and two \textit{vertebral arteries}. The internal carotids provide blood to the \textit{anterior} (front) part of the brain, with one serving the left hemisphere and the other serving the right. The vertebral arteries merge to form the \textit{basilar artery (BA)} \colorSquare{colorScaleBA} near the base of the brain. The basilar artery comes up the brain stem and supplies blood to the posterior (back) part of the brain. The carotid and basilar arteries end at the Circle of Willis.
\datamodel{The ICs and the BA are each a continuous chain of directed, weighted edges where weight can denote width or blood flow.}

\textbf{Circle of Willis:}
The \textit{Circle of Willis (CoW)} is part of the vasculature at the base of the brain and is important for blood distribution.
The CoW regulates blood flow and provides redundant circulation---if part is blocked, blood can still flow to the brain.
At this junction the incoming ICs and BA branch into six cerebral arteries (Fig.~\ref{COW}).
The CoW connects the ICs with the BA through two \textit{posterior communicating arteries (P. Comm.)} \colorSquare{colorScalePcomm}.
It also connects the two anterior cerebral arteries (ACA) through a single artery called the \textit{anterior communicating artery (A. Comm.)} \colorSquare{colorScaleAcomm}.
The communicating arteries allow flow from any input artery to the cerebral arteries and provide a failsafe in case of blockages.
\datamodel{The Circle of Willis is an undirected, weighted cycle.
Connected to the cycle are directed, weighted edges for input and output.}

\textbf{Distributing blood inside the brain:}
The \textit{cerebral arteries} carry blood throughout the brain. Each major artery is named for the region of the brain it supplies.
There are three pairs of cerebral arteries, with one artery in each pair serving the left side of the brain and the other the right. 

The \textit{middle cerebral arteries (MCA)} \colorSquare{colorScaleMCAL} \colorSquare{colorScaleMCAR} branch from the internal carotid arteries (Fig.~\ref{COW}) and serve the middle of the brain. The \textit{posterior cerebral arteries (PCA)} \colorSquare{colorScalePCAL} \colorSquare{colorScalePCAR} branch from the basilar artery (Fig.~\ref{COW}) and serve the back of the brain. The MCAs and PCAs have a directed blood flow stemming from the Circle of Willis.
These arteries all split into two branches at each bifurcation point. \datamodel{The MCAs and PCAs are each directed binary trees.}

The \textit{anterior cerebral arteries (ACA)} \colorSquare{colorScaleACAL} \colorSquare{colorScaleACAR} also branch from the internal carotid arteries (Fig.~\ref{COW}) and serve the front of the brain.
Similar to the MCAs and PCAs, they split into two branches at each bifurcation point and have directed flow from the Circle of Willis into the brain.
However, there exists an undirected bridge between the left and right ACA: the \textit{anterior communicating artery (A. Comm.)} \colorSquare{colorScaleMCAL} \colorSquare{colorScaleAcomm} (Fig.~\ref{COW}).
\datamodel{The ACAs are directed binary trees joined by an undirected edge.}

In Fig.~\ref{Teaser} and Sec.~\ref{goals} we illustrate and describe how this network definition helped us design our novel layout technique and visualization. 

\section{Related Work} \label{sec:relatedwork}

\subsection{Visualizing Cerebral Arteries and Brain Anatomy in 2D} 

As we introduced in Sec.~\ref{sec:domainbg}, conventional 3D cerebrovascular visualizations suffer from issues of occlusion and clutter.
Our goal is to develop a 2D visualization approach that can reduce clutter and assist in visual search tasks.
Previous authors have focused on artery systems in other parts of the body, e.g., Borkin et al.'s heart visualization\cite{6065015}.
To our knowledge no prior work exists on visualizing cerebral arteries as a network. 
However, some authors have examined cerebral arteries using tree visualizations.
E.g., to assist in detecting brain tumors Aydin et al.\cite{Aydin2011} used a tree visualization to represent the density of arteries in a region of the brain. %Dense artery distribution denotes that there is a chance for tumor formation. 
Later Skwerer et al.\cite{skwerer2014tree} used a tree visualization for general analyses.
However, prior work has not visualized the important cyclic Circle of Willis or its feeder arteries.
Prior abstract tree visualizations also have not preserved any spatial position information about the arteries, while knowing exact and relative position is important for diagnosing some cerebral disorders.

Looking beyond arteries, previous authors have used 2D visualizations of other brain structures e.g., for connectomics: the study of neural connectivity in the brain.
Abstract representations have been used extensively to visualize the functional and structural components of the brain (e.g.,\cite{Alper:2013:WGC:2470654.2470724,Worsley913,Mohammed2017,neurolines}).
Alper et al.\cite{Alper:2013:WGC:2470654.2470724} used adjacency matrix and node-link visualizations to study correlations between parts of the brain.
They compared the accuracy of the two approaches and found the node-link visualization to be a more realistic representation of the brain.
However, users faced problems with occlusion and performed worse than using a more abstract adjacency matrix.
Likewise, in Neurolines \cite{neurolines} Al-Awami et al. point out the advantages of using an abstract subway map visualization for nanoscale neuron connectivity.

\takeaway{3D cerebral networks are structurally complicated but are important to understand in connectomics and neuroradiology.
Abstract 2D network visualizations can provide valuable insights while avoiding issues of occlusion, but prior work on cerebral artery visualization has not included key anatomical structures and relative position.}

\begin{figure*}[!tb]
    \includegraphics[width=\textwidth]{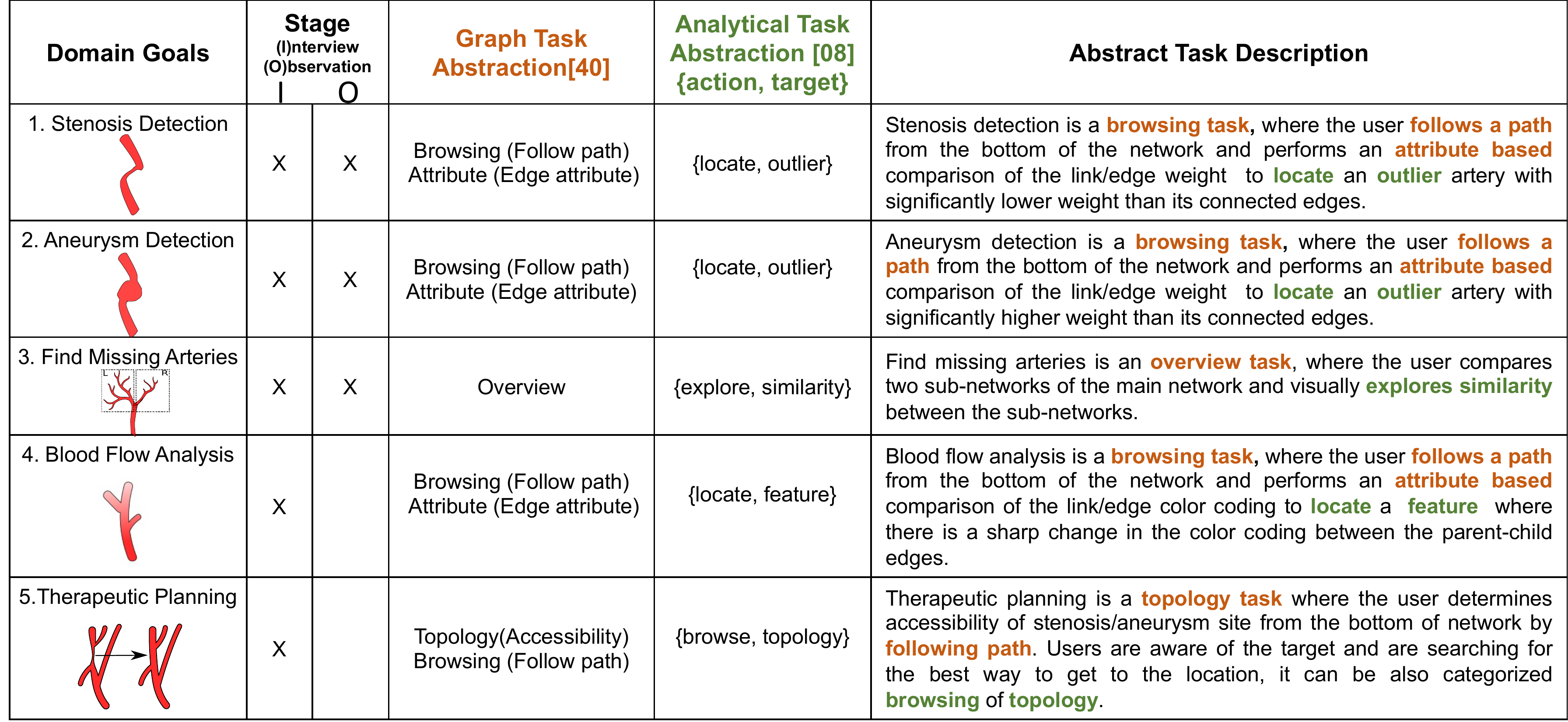}
    \caption{Summary of domain goals with accompanying abstract graph and analytical tasks. }
    \label{domaingoals}
\end{figure*}
\vspace{-0.5 em}

\subsection{Network Visualization} 

\textbf{Readability \& Aesthetics:} The readability of node-link visualizations has been extensively studied, quantified, and summarized (e.g., \cite{battista1998graph,Dunne15Readabilitymetricfeedback,purchase1997aesthetic,Sugiyama02GraphDrawingand,ware2002cognitive}).
Here we discuss three readability criteria which are particularly relevant for cerebral artery analysis: edge crossings, path continuity, and symmetry.
Empirical research emphasizes the fact that a node-link visualization should try to minimize edge crossings\cite{ferrari1969drawing,purchase1997aesthetic} and the idea has been broadly accepted by the community, with multiple algorithms that optimize for fewer edge crossings\cite{Coleman96,davidson1996drawing,eades1989draw,fruchterman1991graph,Sugiyama81}.
Path continuity is an important factor for the path following ability of users. 
Continuous curved paths are more easily perceived than polylines\cite{Kobourov:2015:GPG:2951136.2951162} and
minimizing the number and angle of bends along the path improves path-finding task performance\cite{ware2002cognitive,Kobourov:2015:GPG:2951136.2951162}.
Lipton et al.\cite{lipton1985method} concluded that a good network visualization displays as many symmetries as possible, and symmetry is especially important in our case as we want to enable comparison between the left and right halves of the brain.

\textbf{Tree Layout and Comparison:} While outside of the scope of this paper, an expansive survey of tree visualizations can be found on treevis.net\cite{schulz2011treevis}.
Gomez et al.\cite{guerra2013treeversity,guerra2013visualizing} propose a design space of tree comparison techniques, of which we primarily focus on topology comparison.
Previously Munzner et al.\cite{munzner2003treejuxtaposer} compared the topology of two trees algorithmically.
However, for comparing artery structures between the left and right brain we do not require exact one-to-one matching, nor would it be feasible.
Instead, we use technique similar to that of Holten \& van~Wijk\cite{holten2008visual} which mirrors two tree visualizations in an orientation conducive for comparing hierarchical differences.
\REMOVE{There are two common categories of tree visualizations: space filling techniques inspired by treemaps \cite{shneiderman1998treemaps} and non-space filling techniques for laying out node-link visualizations e.g. Reingold-Tilford\cite{reingold1981tidier}.
While space-filling visualizations are a great way to optimize the usage of screen space to display attribute values, the topology of the tree is not the primary feature encoded.
Therefore, for artery subnetworks that have a tree structure we use a node-link tree visualization and layout.}
\ADD{There are two common categories of tree visualizations: space filling/implicit~\cite{5473227} techniques such as  treemaps~\cite{shneiderman1998treemaps} and non-space filling techniques for laying out node-link visualizations such as Reingold-Tilford~\cite{reingold1981tidier}.
While space-filling visualizations are an effective way to optimize the usage of screen space to display node attribute values, they do not explicitly draw edges~\cite{5473227}.
As discussed in the network data model (see Sec.~\ref{brainvasc}), the human circulatory system is a blood flow network with edges representing vessels.
Therefore, for artery sub-networks that have a tree structure we use node-link visualizations so as to provide edge marks on which we can directly encode blood flow.}

\textbf{Constraint-Based Layout:} Constraint-based layouts impose restrictions on the resulting network spatialization, e.g., to reduce edge crossings\cite{Sugiyama81}, uncover structures or patterns which were not otherwise visible\cite{krzywinski2011hive}, help preserve the user's mental map with interactively changing layouts\cite{he1998constrained}, and impose domain-specific constraints e.g., for biology\cite{barsky2008cerebral,csermely2013structure}. \ADD{ Spatially ordered treemap layouts~\cite{SOT,Nmap, ghoniem2015weighted} take into account the relative spatial position of each node and the distance between the nodes to create a treemap visualization. Spatial ordering reduces the cognitive load associated with finding a node based on its location and supports identification of relationships and trends~\cite{SOT}. }
In the case of cerebral arteries, anatomical and spatial context is crucial for many diagnoses.
Thus we use a novel constraint-based layout to help preserve this context while conserving screen space.
\REMOVE{Generic constraint-based algorithms\cite{schreiber2009generic} and implementations\cite{Colajs} exist but they were overkill for our needs.} \ADD{Generic constraint-based algorithms~\cite{schreiber2009generic} and their implementations~\cite{Colajs} can impose established domain-specific constraints on the network layout. In our case, the network constraints were not well established and emerged only through the design iterations (see Sec.~\ref{goals}). Therefore, we did not use generic constraint-based algorithms~\cite{schreiber2009generic} at the start of our prototype.}

\textbf{Mixed Hierarchy Network Layout:} Networks with both hierarchical and non-hierarchical components---e.g., UML diagrams and cerebral artery systems---are called mixed hierarchy networks. 
Gutwenger et al.\cite{gutwenger2003new} developed a technique for visualizing mixed hierarchy UML diagrams.
To visualize the Circle of Willis we leverage a key feature they propose: that undirected edges should run horizontally while directed edges run vertically and monotonically.
\vspace{-0.5 em}

\takeaway{\REMOVE{Existing node-link network and tree layout techniques are insufficient for visualizing cerebral artery networks.
We use constraint-based and mixed hierarchy layout techniques to ensure key readability criteria are met.} \ADD{Our work builds on existing network and tree layout techniques and adapts it for the context of an arterial network. We use constraint-based and mixed hierarchy layout techniques to ensure key readability criteria are met.}
See Sec.~\ref{sec:CBVisDesign} for more detail.}

\subsection{Abstraction with Context}
\REMOVE{We can simplify a visualization by abstracting away unnecessary complexities of the data. 
E.g., Grabeler et al.\cite{grabler2008automatic} generated tourist maps of selected areas that highlight key landmarks but hide details about streets and locations which are of no interest.
Later Glander et al.\cite{glander2009abstract} abstracted a virtual 3D city model to generate a map of landmarks and used focus+context techniques to view the underlying data.
In both projects the goal was to maximize focus on landmarks by abstracting less relevant data.
This approach has also been commonly used for subway route maps e.g., in Boston, USA\cite{BostonSubway} and Delhi, India\cite{DelhiMetro}.
These subway maps were designed to present routes as clearly as possible using abstraction, but preserved some spatial context such as the relative positions of routes and landmarks.} \ADD{ We can simplify a visualization of spatial data by abstracting away unnecessary complexities of the data. 
For example, Grabeler et al.\cite{grabler2008automatic} generated tourist maps of selected areas that highlight key landmarks but hide details about streets and locations which are of no interest. Often, abstractions supplemented with context can assist the user in understanding the representation~\cite{SOT,yang2016blockwise}. We can provide the context in spatial data by preserving relative spatial position in the visualization. Subway route maps, e.g., in Boston, USA\cite{BostonSubway} and Delhi, India\cite{DelhiMetro}, were designed to present routes as clearly as possible using abstraction but preserved some spatial context such as the relative positions of routes and landmarks. Another method to provide context in an abstract representation is to associate it with the true spatial information. For example, Yang et. al.~\cite{yang2016blockwise} developed an abstract NodeTrix~\cite{henry2007nodetrix} representation to visualize neural connectivity in the brain. In their representation, the NodeTrix visualization was overlaid on a schematic representation of the brain to provide context. In another example, Dykes~\cite{dykes1998cartographic} demonstrated the use of linked views to provide geospatial context with an abstract visualization.}

\takeaway{We use abstraction with context in our design to provide spatial context for the abstract topology visualization of cerebral arteries.}

\section{Domain Goals and Task Abstraction } \label{sec:domgainGoalTask}
In order to better understand the diagnostic tasks (domain goals) and workflow in neuroradiology for cerebrovascular diseases, we worked closely with a neuroradiologist at the Brigham and Women's Hospital (24 years of experience) and a neurological MRI physicist at the Dana-Farber Cancer Institute (10 years of experience). Both experts are co-authors of this paper. In order to generate the domain goals we conducted a series of interviews with these experts, solicited their feedback in our iterative design process, and further validated the curated domain goals through a series of observational studies.

\textbf{Expert Interviews:} We conducted a series of 10 open-ended conversational style interviews with the 2 experts which took on average 90 minutes each. The open-ended style gave our experts the freedom to provide us with sufficient introduction and supporting information from the field of neuroradiology, discuss their current diagnostic practices, and share the pros and cons of the existing techniques. 

\textbf{Expert Observation:} We conducted 4 observational studies with the 2 experts at Brigham \& Women's Hospital with each observation session lasting on average 6 hours. For the studies we followed a shadowing methodology, a common observational procedure for premed students to understand a physician's typical work activities in a clinic or hospital setting\cite{rothwell2012junior}. In requirement analysis literature\cite{goguen1993techniques} shadowing resembles a protocol analysis procedure. In protocol analysis, the analyst observes experts in their natural workflow and the experts talk out loud and explain their tasks. This protocol analysis procedure worked well with our experts' workflow as the expert already narrates the diagnostic process aloud, transcribed in realtime with dictation software, for their required radiological case study report. Shadowing importantly enabled us to uncover domain goals that were not shared by the expert during the interview procedure.

\subsection{Domain Goals}

Based on hand-written notes collected during the expert interviews and observational studies, we were able to apply open-coding and summarization techniques to identify the expert's domain goals for diagnosis of cerebrovascular diseases in medical images. All domain goals are applicable to diagnoses with either CTA or MRA imaging datasets. In ranked order of importance, the observed domain goals are:

\textbf{1. Stenosis Detection}: Identify an intracranial stenosis, i.e., narrowing of an artery inside the brain. Over time, fat can be deposited along the walls of medium and large arteries in the body, causing them to become narrowed or even blocked. To detect a stenosis experts use artery visualization to look for abnormally narrow arteries. 

\textbf{2. Aneurysm Detection}: Detect a brain aneurysm, which is an abnormal widening or ballooning of a cerebral vessel. Brain aneurysms occur when an injury, congenital disability, or other diseases weakens the wall of the vessel. To detect aneurysms experts use artery visualizations and look for the balloon-like bulge of artery walls. The doctors additionally examine the 2D source image `slices' to look for potential bleeding in the brain, which is not detectable in artery visualizations.

\textbf{3. Find Missing Arteries}: The distribution of arteries in the left and the right hemisphere of the brain should be fairly symmetrical for a healthy patient. A highly uneven distribution of arteries in the left and right hemisphere may be due to a blockage (clot) which obstructs blood flow. These structural differences between the left and right sides (vascular malformations) can be indicative of this artery occlusion, a common indicator of stroke \cite{tanaka2008chronic}. If no blood is flowing in the arteries, then it will not be visible in the images thus doctors look for ``empty'' or missing branches in the artery visualization. 

\textbf{4. Blood Flow Visualization}: Blood flow volume data is very useful for the detection of cerebrovascular diseases since major vascular abnormalities are accompanied by a disruption of blood flow. For the detection of these diseases, a blood flow visualization should enable the differentiation between regions of regular and irregular blood flow. Currently this data, either calculated through a hemodynamic blood flow simulation or interpolated from the luminosity of the contrast dye, is not readily available to radiologists in their clinical imaging suites. Experts sometimes try to qualitatively interpolate this information from the luminosity of voxels in the artery visualization or raw imaging data.

\textbf{5. Therapeutic Planning}: To treat cerebrovascular diseases, interventional radiologists and surgeons repair diseased arteries through invasive methods including stent insertion or balloon angiography in order to widen the artery and restore blood flow. 
These interventional procedures require careful planning and execution. An essential part of interventional surgical planning is the ability to carefully navigate the surgical equipment to the site of stenosis and aneurysm in the arteries, guided through medical images and artery visualizations of the patient.

\vspace{-0.5 em}

\subsection{Task Abstraction} 
As discussed in the preceding section, the primary neuroradiology domain goals involve finding abnormalities in the network of cerebral arteries. Following conventional design study and nested model procedures\cite{5290695,sedlmair2012design}, we frame the domain tasks as low-level analytic tasks both for generalizability as well as for an aid in visual encoding design choices. In this paper we use utilize the Brehmer and Munzner multi-level task typology\cite{Brehmer2013} to determine the \textit{\{Action, Target\}}. As we established in Sec.~\ref{brainvasc}, the cerebral arteries can be defined as a network. Here we contribute a network task abstraction of the neuroradiology domain goals using the Lee et al. graph task taxonomy\cite{lee2006task}. For example: \textit{``Stenosis detection is a browsing task, where the user follows a path from the bottom of the network and performs an attribute based comparison of the link/edge weight.''} These network task taxonomy classifications were essential for developing our novel network layout for the CerebroVis. The domain goals, network tasks, and general analytical tasks are summarized in descending order of importance for a diagnosis in Fig.~\ref{domaingoals}. Both the general abstract and network task definitions helped ensure that the final tool and layout supported the domain goals.

\section{CerebroVis Design Process} \label{goals} 

\begin{figure*}[tb]
    \includegraphics[width=\textwidth]{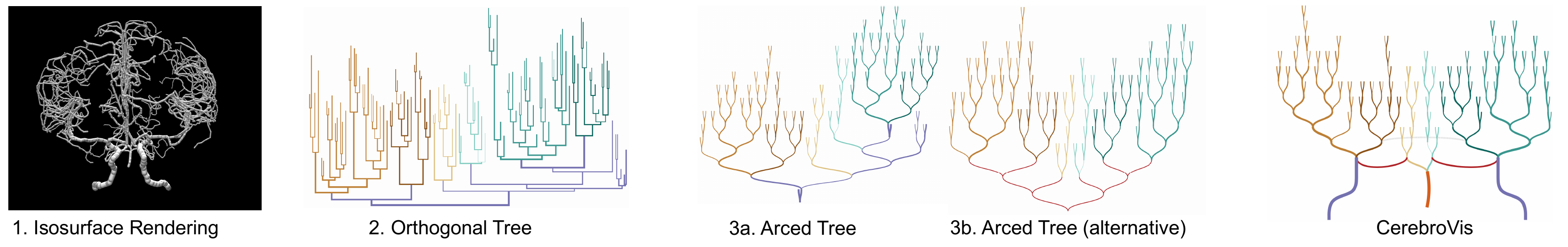}
    \caption{The visual design evolution of CerebroVis from (left) 3D rendering to (right) final 2D network representation.} \label{designIteration}
\end{figure*}

We developed CerebroVis in a user-centered iterative design process with domain experts in order to ensure validity, accuracy, and applicability of our final design. We also relied on our abstract and network task analyses to ensure appropriate choice and visual design of our final 2D encoding. In the following sections we discuss our design iterations (see Fig.~\ref{designIteration}) and summarize the design requirements in terms of both visual encoding language as well as network topology.

\subsection{Iterative Design Process and Goals Formulation} \label{subsec:designiteration}

\textbf{Iteration 1---3D  Rendering:}
In our first design iteration we focused on the development of a clean and clear 3D rendering, close to conventional approaches. We used Sharkviewer \cite{SWCv}, an existing open source tool, to create a 3D  visualization of the cerebral artery structure (Fig.~\ref{designIteration} (1)). Although conventional techniques use both volume rendering and isosurface techniques, in both cases the goal is to produce a rendering with clear hard surfaces and structure. The rendering is constructed using the spatial information provided in the digital segmentation data (Sec.~\ref{sec:DataDescription}). The visualization is interactive and the experts can pan, zoom, and rotate. This rendering was also used in interviews to discuss the existing diagnostic techniques with experts (Sec.~\ref{sec:relatedwork}).

\textit{Expert feedback:} The experts like the 3D rendering as it preserves the spatial and anatomical properties of the arteries. However, in order to maximize efficiency and minimize cognitive load, the experts prefer views with minimal or no interaction. Most importantly, the experts noted that the 3D rendering suffers from issues of occlusion and is liable to inaccurately render small merging or tangled geometries (i.e., small or tangled features are merged into single 3D feature). The 3D rendering also proved difficult to use with diagnostic tasks that require ``path following'' to trace a particular artery through the 3D space.

\textbf{Iteration 2---2D Tree Diagram:} In order to counter the issues of occlusion and rendering artifacts, support path following tasks, and eliminate interaction with the visualization, we created an abstract 2D orthogonal tree diagram visualization of the arteries. To make the pseudo-hierarchical artery structure readable, we imposed a binary tree structure on the network. Also, in order to preserve some spatial components of the artery geometry in support of stenosis and aneurysm detection tasks, we encoded artery width and length in the edge width and length.

\textit{Expert feedback:} Although the 2D tree diagram eliminated the need for interaction, it had several major drawbacks including the lack of apparent symmetry preservation between the left and right hemispheres, overemphasis on (normal/healthy) artery sub-trees with very long branches, and the inability to easily compare the widths of arteries due to the varying lengths. This last point in particular led to the experts inability to identify stenoses, aneurysms, and missing arteries.

\textbf{Iteration 3---2D Tree Diagram without Artery Length Encoding:} In the third design iteration we made two major changes to the visual encoding: a new tree rendering style and a new tree layout. For the tree drawing, we moved from an orthogonal to arced tree (Fig.~\ref{designIteration}, Iteration 3a), which was laid out using the Reingold-Tilford Algorithm\cite{reingold1981tidier}. For consistency, we use the implementation provided by the D3.js library\cite{Bostock:2011:DDD:2068462.2068631}. Also, in order to better support comparison of artery widths for diagnosis of stenosis and aneurysm detection we removed the artery length encoding. In addition, the arced style more closely resembles the real physical appearance of arteries as compared to the sharp right-angles of the orthogonal layout and is easier for path following as discussed in Sec.~\ref{sec:relatedwork}. Finally, we developed an alternative arrangement of the arteries (Fig.~\ref{designIteration}, Iteration 3b) specifically designed to allow visual comparison of the symmetry between the left and right sides of the brain. In this view each tree depth is at the same vertical level, and in the horizontal arrangement branches are pivoted to encode some aspects of spatial position on the left or right side of the brain.

\textit{Expert feedback:} The new arced style for the tree diagram was much appreciated for its clarity, easy path following, and ability to compare artery widths. The new layout also more accurately presented the balance between the left and right hemispheres. However, the tree representation did not accurately present the network data near the base of the system, in particular the Circle of Willis (CoW, see Sec.~\ref{brainvasc}). The CoW is a key anatomical feature of the cerebral artery system and also provides an important spatial point of reference for overall layout interpretation.

\subsection{CerebroVis Design Requirements} \label{subsec:designrequirement}

Based on our task abstraction (Fig.~\ref{domaingoals}) and expert feedback from our design iterations (Sec.~\ref{subsec:designiteration}), we developed several design requirements to inform our final design:

\textbf{DR1.} \textit{Preservation of expert mental model:} Through our iterative design we established that certain vascular structures, spatial layouts, and visual cues are necessary to provide sufficient context for accurate interpretation. Purchase et al.\cite{purchase2006important} found that as a user observes and understands the layout of a graph, she creates an internal representation of the information about the data as conveyed in visual forms. For a neuroradiologists, identification of the arteries of the Circle of Willis (CoW), along with its connected arteries' shapes and sizes, is essential. Therefore, to provide sufficient anatomical context we divide this design requirement further into:

a. Distinct and consistent representation of the CoW.

b. Preserve position of cerebral arteries relatively to the CoW.

c. Preserve natural variability within the cerebral arteries.

These design requirements allow users to read and interpret the network visualization.

 \smallskip 
 
\textbf{DR2.} \textit{Highlight abnormalities:} Identification of abnormal topology, geometry, and attributes of the cerebral arteries is the primary domain goal as established by our task analysis. These diagnostic tasks heavily informed the iterative design as discussed in Sec.~\ref{subsec:designiteration}. The specific design requirements to support the identification of these abnormalities include:

a. Provide a readable network visualization of the cerebral network.

b. Display abnormal narrowing or widening of the arteries.

c. Compare topology between left and right cerebral arteries.

d. Show direction and volume of blood flow within the arteries.

These design requirements support abnormality detection domain goals (Fig.~\ref{domaingoals} (1--4)).

 \smallskip

\textbf{DR3.} \textit{Help experts gain confidence with interpretating an abstract encoding:} Once an expert has identified an abnormality indicative of disease in the cerebral arteries, disease treatment and intervention procedures need to be determined and executed. In this follow-up step an expert needs to identify the abnormality in the 3D rendering as well as original imaging data. In order to make the network layout interpretable with sufficient context this design requirement consists of:

a. Allow examination of abnormal geometry.

b. Locate the abnormality in the 3D rendering.

Ability to locate and examine an abnormality plays a critical role in therapeutic planning procedure (Fig.~\ref{domaingoals} (5)).

%  \smallskip 

\REMOVE{With these summarized design requirements we completed a final design iteration and created our new solution CerebroVis. The final network visualization and supporting tool are discussed in the following section.}

\section{CerebroVis} \label{sec:CBVisDesign}

\subsection{Spatially Contextualized Network Layout} \label{NetworkLayoutDesign}
As illustrated in Fig.~\ref{Teaser}, the cerebral network consists of two subnetworks: the Circle of Willis (CoW) (an undirected weighted cycle) and the directed binary trees of the cerebral arteries. In the following sections we discuss the layout and design of each subnetwork separately.

\begin{figure}[tb]
    \includegraphics[width=\linewidth]{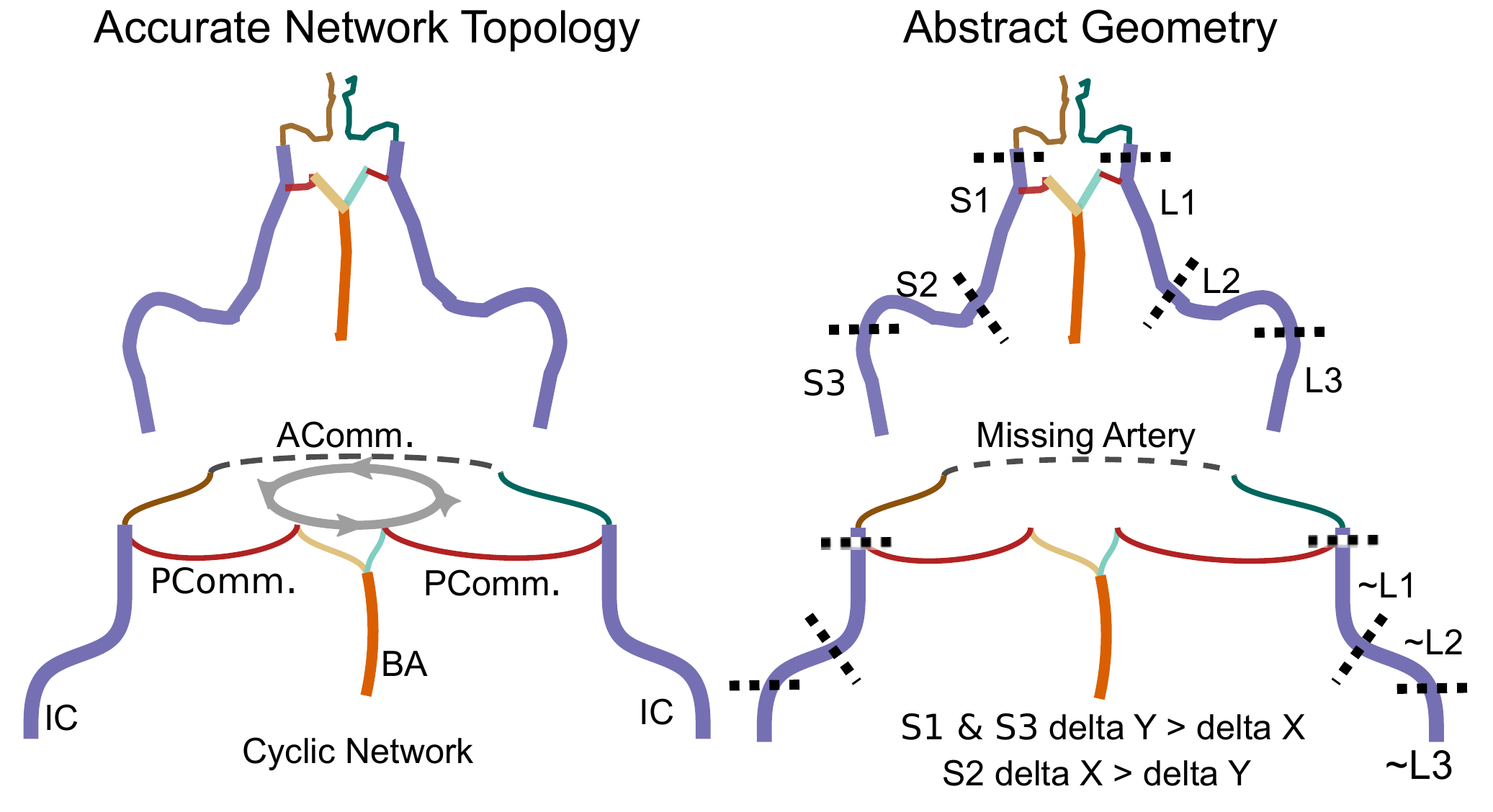}
    \caption{
    Left: Reconstructing of the Circle of Willis (CoW) cycle of the cerebral artery network.
    Right: How CerebroVis abstracts the carotid artery geometry to preserve a frame of reference.} \label{cowconversion}
\end{figure}

\textbf{Circle of Willis (CoW):} The CoW is an undirected weighted cyclic network composed of the P. Comm., A. Comm., Internal Carotids (IC), Basilar Artery (BA) and parts of Anterior Cerebral Artery (ACA) (Sec.~\ref{brainvasc}). To represent the accurate cyclic structure of the CoW we arc the A. Comm. upward and the P. Comm. arteries downward as shown in Fig.~\ref{cowconversion} (Accurate Network Topology). Any missing arteries in the CoW are represented using a dashed line. In our data\cite{wright2013digital} the A. Comm. was excluded to maintain a binary tree structure thus a dashed line is used to represent the artery (see Fig.~\ref{cowconversion}). 

To match the internal representation of the cerebral network and preserve the user's mental model, we abstract the geometry of the IC and BA (see Fig.~\ref{cowconversion}: Abstract Geometry). 
Abstraction of the geometry of the carotid arteries is a two step process: we first find the length and the total number of bends. To estimate the bends we use the spatial position of each segment that makes up the IC.
In Fig.~\ref{cowconversion} (Abstract Geometry), there are a total of 3 bends. The bends are broken down as S1--S3 in the Fig.~\ref{cowconversion} for explanation. In S1 and S3 the artery moves in a vertical direction, therefore a change in the y vertical direction is more than change in the x direction. Next, a Beizer curve is generated with an equal number of bends and the length of the curve is proportional to length of S1--S3 in the carotid arteries. Additionally, the width of each artery curve is proportional to the width of the artery width in the original data. \textbf{\underline{Requirement Satisfied: \textit{DR1a}}}---We distinctly and consistently represent the cyclic network structure of CoW and the shape of the carotid and basilar arteries.

\textbf{Global Spatial Position of Cerebral Arteries:} Global spatial preservation operates at two levels. The first level divides the left and right hemispheres of the brain. This level ensures that cerebral artery trees on the left and right never cross over, which would be anatomically inaccurate. Within each hemisphere the arteries are located in specific positions. For example, from the center moving outward we place the PCA, ACA, then MCA (a schematic of this is presented in Fig.~\ref{Teaser}: Spatial Constraints). This positioning scheme also ensures that cerebral trees maintain familiar position with respect to the CoW. 
% In CerebroVis, the internal carotid arteries branch out as MCA and ACA, and the central basilar artery branches out to the PCA. 
\textbf{\underline{Requirement Satisfied: \textit{DR1b}}}---Spatial constraints restrict the placement of the cerebral artery trees to a deterministic area of the display and preserve position relative to the CoW. Our validation (Sec.~\ref{sec:relatedwork}) confirms that this design enables experts to distinguish between cerebral trees. 

\begin{figure}[tb]
    \centering
    \includegraphics[width=.8\linewidth]{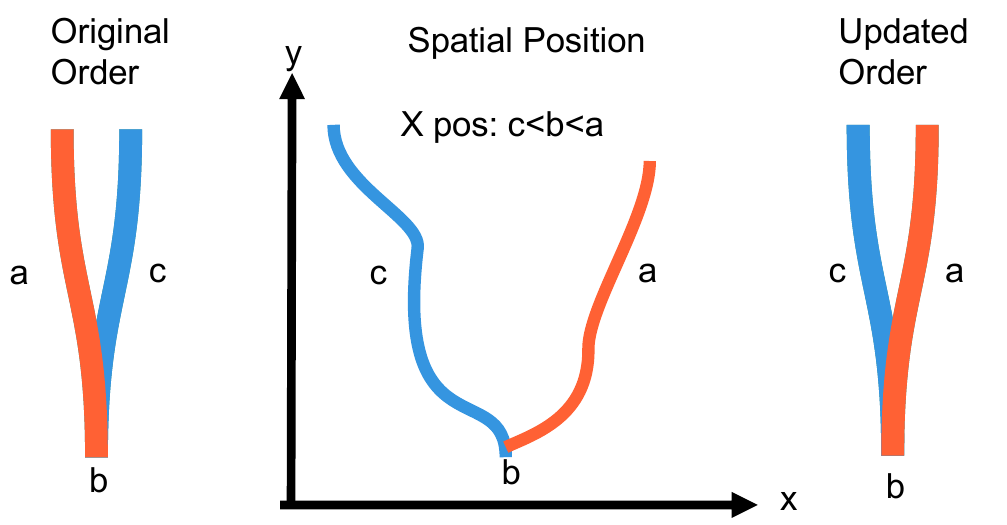}
    \caption{At each artery bifurcation CerebroVis preserves the relative spatial context of each subtree by comparing average horizontal position.} \label{fig:orderarteries}
\end{figure}

\textbf{Local Spatial Position of Cerebral Arteries:} Local spatial preservation is tied to the aspect of safeguarding spatial context within each cerebral tree (PCA, MCA, and ACA). Each artery in the original data exists as a 3D geometry.
In CerebroVis, we preserve approximate spatial (x,y) positions for each artery of the cerebral trees.
Vertical height $y$ of each artery is represented by the height/depth of the edge in the cerebral tree.
The $x$ position of arteries is preserved relative to their branching site.
For example, in Fig.~\ref{fig:orderarteries} (Original Order) the original artery order from the source data is presented.
If the original order does not match the the relative spatial position of any two arteries, as is the case in Fig.~\ref{fig:orderarteries} (Spatial Position) where artery $c$ is to the left of artery $a$, the ordering is swapped (Fig.~\ref{fig:orderarteries} (Updated Order)).
This process is repeated from leaf nodes to the root which ensures that the arteries closer to the CoW are positioned closer to the CoW and otherwise away from the CoW on the $x$ axis. \textbf{\underline{Requirement Satisfied: \textit{DR1c}}}---Local position preservation ensures that when experts look at an edge in the network they have an approximate sense of the position of the artery in the brain.

\begin{figure}[tb]
    \includegraphics[width=\linewidth]{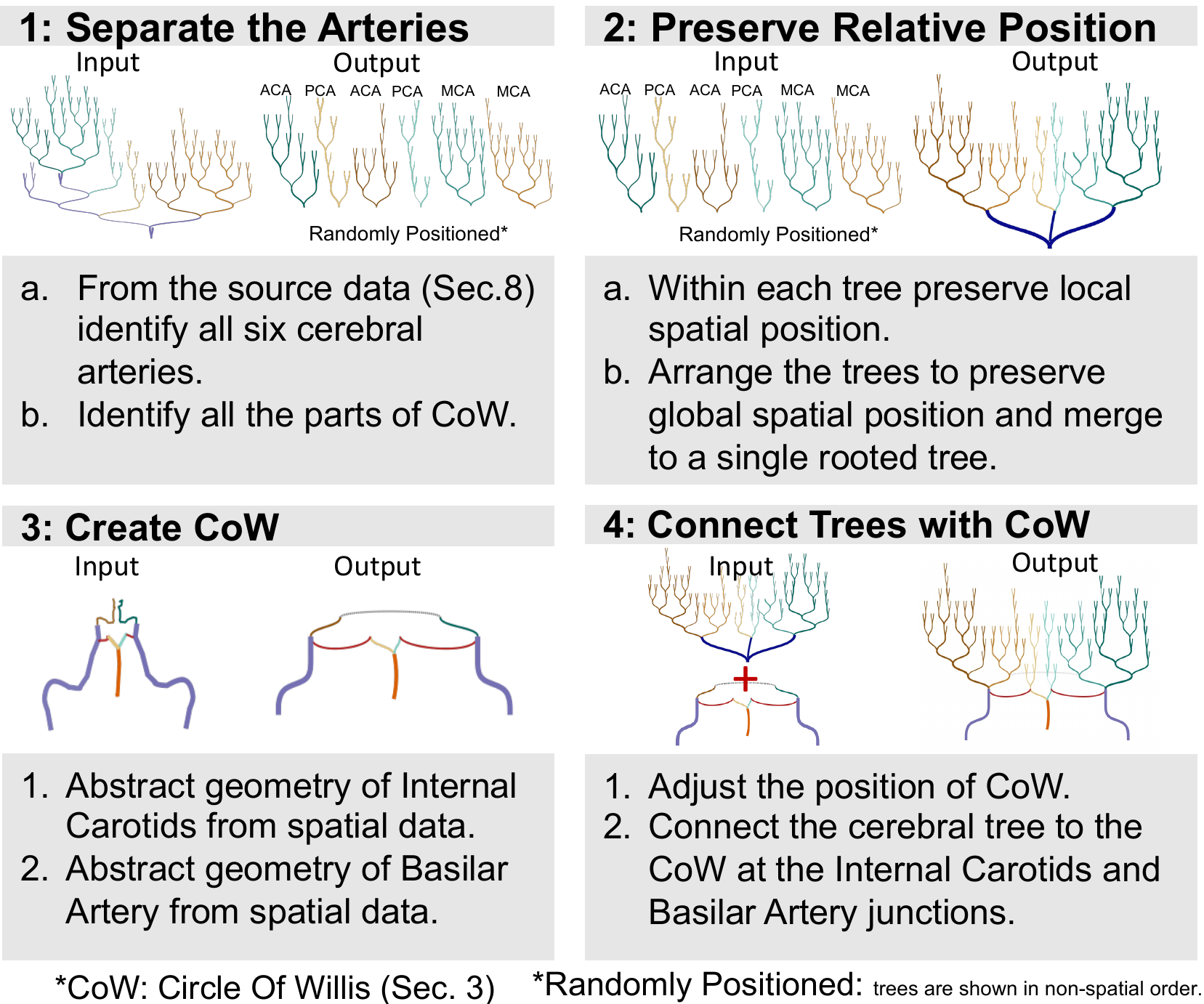}
    \caption{
    CerebroVis converts an input of a single rooted tree into a mixed hierarchy network visualization via four steps.} \label{fig:algorithm}
\end{figure}

\textbf{Layout Technique:}
Our novel layout technique CerebroVis, outlined in Fig.~\ref{fig:algorithm}, uses the data discussed in the data section (Sec.~\ref{sec:DataDescription}) to create the representation.
\REMOVE{CerebroVis uses an orthogonal arrangement as opposed to a radial arrangement.
The orthogonal arrangement of the network aids faster lookup of hierarchy\cite{burch2011evaluation} and supports the experts' mental model of viewing the brain from the perspective of orthogonal faces of the data.}
\ADD{CerebroVis uses a layered upward planar drawing by DiBattista et al.~\cite{battista1998graph} which enables a fast lookup of the hierarchy~\cite{burch2011evaluation}, eases topology comparison, and supports the experts' mental model of viewing the brain from the perspective of orthogonal faces of the data.}
The implementation of the technique  is available at https://osf.io/e5sxt/

\subsection{Visual Encoding to Highlight Abnormalities}
Based on the domain goal analysis (Sec.~\ref{sec:domgainGoalTask}) and iterative design goal formulation (Sec.~\ref{subsec:designiteration}), CerebroVis needs to provide an easily readable network visualization with a visual encoding suitable for the following tasks: (1) locate outlier artery shape (stenosis and aneurysm), (2) explore similarity to find missing arteries, and (3) locate a group of arteries with abnormally low blood flow.

\textbf{Readable Network Visualization:} In support of our design goals, CerebroVis presents a readable network layout inspired by literature in network visualization (Sec.~\ref{sec:relatedwork}). The final layout (Fig.~\ref{Teaser}) minimizes edge overlap, replaces long curved edges with small arced edges, and distinguishes between direction (uni-directional and bi-directional) network edges as explained in the network layout section (Sec.~\ref{NetworkLayoutDesign}). The reduced diversion due to branch disentanglement represents accurate network topology and reduces the cognitive load for path following in the cerebral network. Minimized edge overlaps enable the occlusion-free inspection of each artery. \textbf{\underline{Requirement Satisfied: \textit{DR2a}}}---We provide a readable visualization of the cerebral network. 

\begin{figure*}[tb]
    \includegraphics[width=\textwidth]{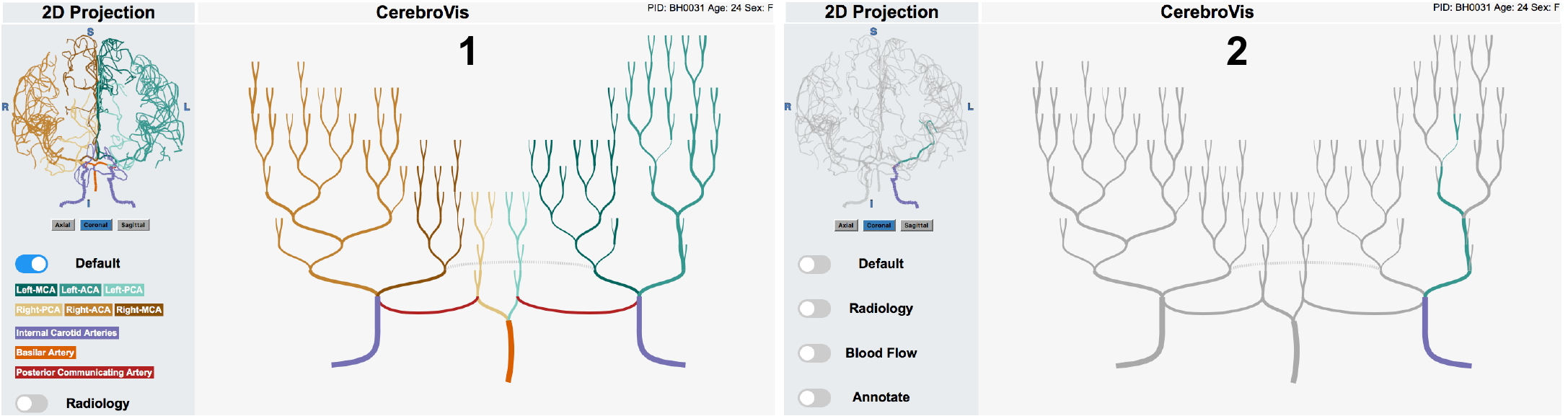}
    \caption{
    The CerebroVis Dashboard with categorical coloring to differentiate arteries.
    Left: A cerebral artery scan with a stenosis in the MCA~\colorSquare{colorScaleMCAR}.
    Right: Users can click on an artery mark in CerebroVis and the corresponding mark is highlighted in the 2D projection.
    This feature allows users to validate the stenosis with the underlying geometry and plan for therapeutic surgery.} \label{fig:workflowexample}
    
\end{figure*}

\begin{figure}[tb]
    \includegraphics[width=\linewidth]{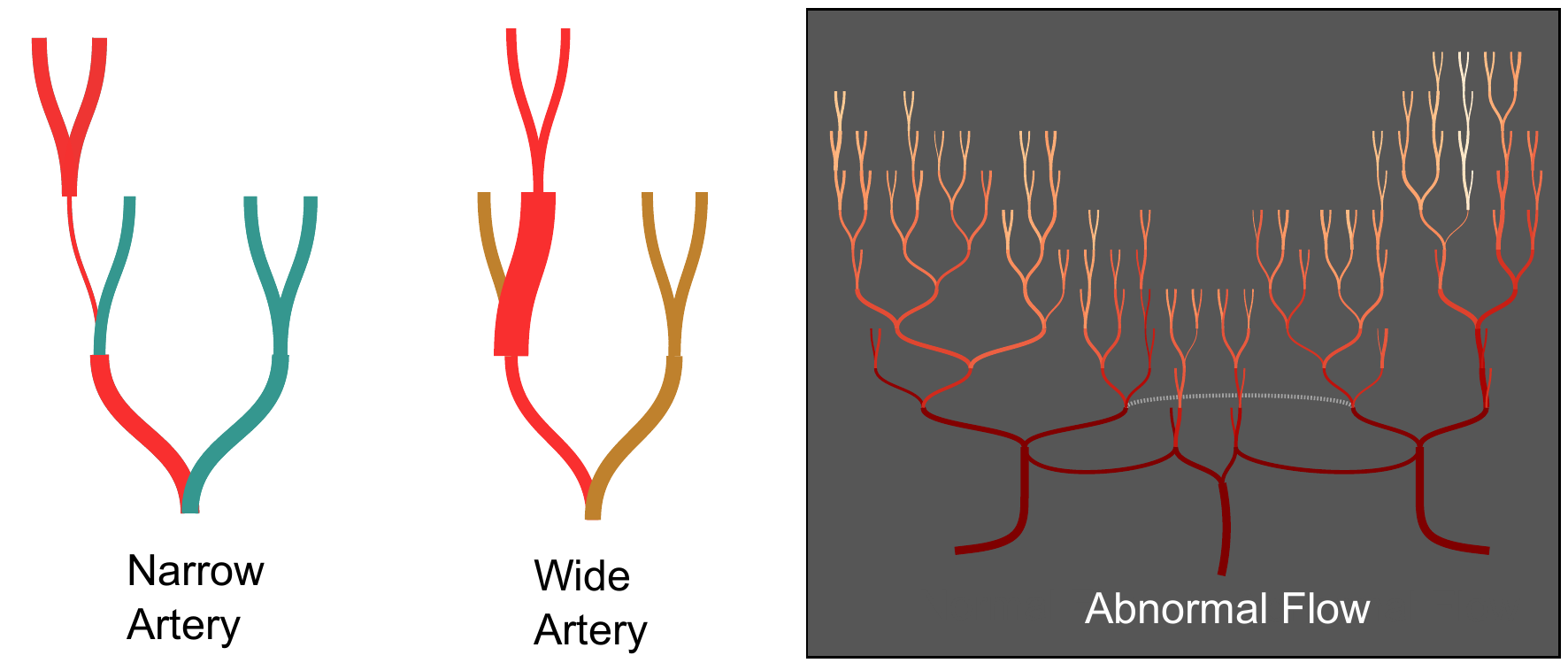}
    \caption{
    Left: Examples of an abnormally narrow artery (``stenosis'') and wide artery (``aneurysm''), with the relevant branches colored red \colorSquare{colorScaleRed}.
    Right: Example of a blood flow color encoding with a blockage disrupting normal flow of blood with blood flow colored on a scale between \colorSquareBorder{colorScaleMin} and \colorSquare{colorScaleMax}.}
    \label{fig:stenosisblood}
\end{figure}

\textbf{Locate Outlier Artery Width:} CerebroVis visually encodes the width of the artery. Network edges are scaled proportional to the average width of the representative artery. Arteries linearly taper (narrow) as they move away from the Circle Of Willis. Thus we use a linear scale to map the width of the arteries to network edge weights. Whenever an artery significantly deviates from the natural linear tapering and appears abnormally narrow or wider than its parent and child arteries (Fig.~\ref{fig:stenosisblood}), those arteries can be potential sites of stenosis and aneurysm in the cerebral network. \textbf{\underline{Requirement Satisfied: \textit{DR2b}}}---We display abnormal narrowing or widening of the arteries.

\textbf{Explore Similarity:} Topology comparison of the cerebral arteries is an overview task for the neuroradiologists (Sec.~\ref{sec:domgainGoalTask}). To compare topologies, experts estimate depth and width difference between the left and right arteries (PCA, ACA, and MCA). CerebroVis enables depth comparison between left and right arteries by placement of branching sites at the same vertical height. For example, in Fig.~\ref{fig:workflowexample} (1) the MCA on the right~\colorSquare{colorScaleMCAR} and the left~\colorSquare{colorScaleMCAL} begin at the same vertical height. Additionally, each subsequent branching site is positioned at the same vertical height. Therefore, the difference between the left and right trees can be determined through comparison of the vertical positions of leaf arteries. The comparison of width is possible at an overview level as the arteries take space proportional to their tree width. For subtle differences, at each bifurcation site the length of an artery roughly encodes the width of the subtree at that bifurcation. For instance, in Fig.~\ref{fig:workflowexample} (1), the MCA on the left~\colorSquare{colorScaleMCAL} has a greater width than the MCA on the right~\colorSquare{colorScaleMCAR}. \textbf{\underline{Requirement Satisfied: \textit{DR2c}}}---We enable comparing topology between left and right cerebral arteries.
 
\textbf{Analyze Blood Flow:} In CerebroVis an edge also encodes the amount of blood flow in an artery. To encode the flow, we use a linear color scale between \colorSquareBorder{colorScaleMin} and \colorSquare{colorScaleMax} (see Fig.~\ref{fig:stenosisblood}).
To test the visual output of the encoding, we simulate blood flow in the cerebral arteries with a simple linear model. The model divides a fixed amount of blood flow through the arteries, where the flow in an artery is proportional to its width and inversely proportional to the height of the artery from CoW. The model is discussed in the Supplemental Material. The output from the blood flow volume model with a simulated thrombosis or blockage in an artery, shown in Fig.~\ref{fig:stenosisblood} (Abnormal Flow), demonstrates how the same set of arteries will show a sudden white outlier indicating a blockage. We believe complex blood flow models can also be represented with CerebroVis, and we envision testing more models in future. \textbf{\underline{ Requirement Satisfied: \textit{DR2d}}}---We show the direction and volume of blood flow within the arteries.

\subsection{Linked Views to Maintain User Confidence}

CerebroVis preserves context with its spatially constrained layout. However, it does not preserve the true spatial position and 3D geometry of an artery. To provide anatomical context, we created an accompanying interactive dashboard which uses a linked view approach between edges of CerebroVis and a 2D projection of a 3D visualization (Fig.~\ref{fig:workflowexample}). The projection can be changed to show arteries in the conventional radiology views from the front, top, and side of the head. For linking, a user is able to click on any edge in the network and the corresponding artery in the 3D projection is highlighted (Fig.~\ref{fig:workflowexample} (2)). The highlighted 2D projection artery can be used to validate the geometry and location of an artery in the brain.

To identify and distinguish cerebral arteries easily in the 2D projection and CerebroVis network visualization, we use a categorical color map. This color coding uses the same hue for arteries on the same side of the brain, but varies the saturation of the color. The saturation captures the depth perspective of the arteries. For example, ACA~\colorSquare{colorScaleACAR} provides blood to the front of the brain so they are darkest and PCA~\colorSquare{colorScalePCAR} provides blood to the back of the brain and they are lightest. In the CoW each type of artery receives a unique color, with a bright red hue for P. Comm.~\colorSquare{colorScalePcomm}, to highlight it in the 2D projection.

\subsection{CerebroVis Web Application}
An open source web-based implementation of CerebroVis (Fig.~\ref{fig:workflowexample}) is available at \href{https://aditeyapandey.github.io/CerebroVisProject/}{aditeyapandey.github.io/CerebroVisProject}
We use D3.js\cite{Bostock:2011:DDD:2068462.2068631} for the implementation of CerebroVis.
Additional details about the tool are available in the Supplemental Material.

\section{Data and Implementation Robustness}
\label{sec:DataDescription}

For the design, development, and evaluation of CerebroVis and the new 2D layout we used a collection of open source MRA datasets of 61 healthy patients from Wright et al.\cite{wright2013digital}.
We chose this collection of data as it is representative of real clinical data (i.e., ecological validity) both in terms of imaging modality (MRA) and quality.
The data is also freely available and has had all identifiable patient information properly removed.
Each dataset additionally includes the 3D cerebral artery geometry extracted from the MRA scan using a conventional medical imaging segmentation technique\cite{despotovic2015mri}.
The segmentation extracts a binary tree stemming from the Basilar Artery (BA).
Starting from the BA, all visible connecting vessels are added with the exception of the anterior communicating arteries in order to maintain a binary tree structure.

Our data ingestion process is discussed in detail in the Supplemental Material but we provide an overview here.
The vascular structure data is provided as an swc tabular file---a format originally developed for neural connectivity networks\cite{SWCv}.
In the file each artery is represented by a chain of artery segments, each of which is described by its position, size, and parent/child edge relations.
We convert the swc data into a rooted tree to show the hierarchical structure of the cerebral arteries which we can then modify to create a mixed network that includes the Circle of Willis cycle.

To validate our data conversion pipeline, network layout technique, and visualization design we loaded and visualized the 61 scans from Wright et al.\cite{wright2013digital}.
We manually assessed the robustness of the conversion according to three criteria:
\REMOVE{
(1) \textit{No Missing Arteries in the CoW (Circle of Willis)}: In all 61 scans we verified the presence and correctness of the CoW arteries with the help of Linked Views.
(2) \textit{Non-Binary Bifurcation is Displayed}: Our technique is designed to detect and visualize non-binary bifurcation of the cerebral arteries, the Internal Carotid Arteries, and the Basilar Artery.
We did not find non-binary bifurcations in any of the 61 scans.
(3) \textit{Views are Linked Properly}: We randomly selected 20 scans, for which we traversed the entire cerebral artery system using the CerebroVis Linked Views.
In all cases artery marks in the 3D projection were linked properly to their counterparts in the 2D projection.}
\ADD{
(1) \textit{Visual Similarity of the CoW (Circle of Willis)}:
In each scan we verified the similarity of the shape of CoW in the 3D projection and CerebroVis. 
(2) \textit{Spatial Constraints are Imposed Properly}:
In each scan we examined the position of the cerebral artery branches relative to the CoW then randomly chose one branch to traverse and validate artery positions using the linked views.
(3) \textit{Views are Linked Accurately}:
We randomly selected 20 scans, for which we interactively traversed the entire cerebral artery system using the CerebroVis Linked Views.
In all cases artery marks in the 3D projection were linked properly to their counterparts in the 2D projection.
\textbf{Rationale}: Criteria 1 and 2 validate our primary design goals (discussed in Sec.~\ref{subsec:designrequirement}).
Criteria 3 validates the accuracy of data binding between the 3D projection and CerebroVis visualization.
}

\section{Comparative Evaluation}

To validate and evaluate the design and functionality of CerebroVis against a conventional 3D  visualization we conducted a mixed-methods study\cite{johnson2004mixed}.
The evaluation included a controlled task-based experiment where experts diagnosed a simulated intracranial artery stenosis (narrowing of the artery) and a semi-structured interview.

\subsection{Participants}

For our study we recruited three neuroradiologists from Brigham \& Women's hospital who had no prior experience with CerebroVis.
They had 8, 25, and 40 years of experience.
In addition to their participation in our evaluation, they provided additional post-experiment feedback on this research and thus are included as co-authors on this paper.

\subsection{Methodology}

We arranged an evaluation session with each participant.
Each session lasted at minimum an hour, with some of longer duration based on expert availability for an extended interview to solicit additional qualitative feedback.
Participants spent \textasciitilde 15 mins on the introduction and tutorial, but as the tutorial was interactive it often lasted longer than planned.
The next \textasciitilde 25--30 mins were spent on a within-subjects task-based controlled experiment which simulated intracranial artery stenosis diagnosis tasks.
We chose stenosis diagnosis because it is a primary cause of cerebrovascular disorders\cite{ischemicstroke}.
Finally we performed \textasciitilde 15 mins of semi-structured interview to solicit qualitative feedback on features, usability scenarios, drawbacks and disadvantages, and potential directions for future work.

\subsection{Stimuli and Tasks}

We used as our stimuli two black-and-white cerebral artery visualizations: a 3D  visualization (3D) (Fig.~\ref{visualizationtech} (C)) and the CerebroVis (CV) 2D network (Fig.~\ref{visualizationtech} (D)).
\REMOVE{We chose a black-and-white color encoding to emulate conventional visualizations used clinically. 
As our data collection was of presumably healthy patients (Sec.~\ref{sec:DataDescription}), under the close guidance of our domain collaborators we simulated stenoses in 5 random scans by manually introducing an abnormal artery narrowing into the data.
The altered scans were then randomly mixed with 5 unaltered scans.} 
\ADD{We chose a black-and-white color encoding to emulate the conventional clinical visualizations.
The 3D isosurface visualization similarly echoed the clinical convention of limited interaction with only clockwise/counterclockwise rotation of variable speed around the vertical axis enabled (i.e., no panning, zooming, or free-form rotation). To control the rotation, we provide options including stop the rotation, change the default direction of rotation, and alter the speed of rotation. In the 2D network representation, users were only shown the static novel 2D CerebroVis layout without the linked view or context. As our data collection was of presumably healthy patients (Sec.~\ref{sec:DataDescription}), we simulated stenoses in 5 random data sets under the close guidance of domain collaborators. All of the injected stenoses were severe with over 70\% narrowing of the artery. We induced only severe cases because they pose a greater risk to life and are the top priority for clinical diagnosis. The location of the stenoses were randomly determined but were restricted to the cerebral branches PCA, MCA, and ACA. The altered scans were then randomly mixed with 5 unaltered scans. }
Each participant was asked to read the 10 scans in a randomized order with one visualization, then switched to the other visualization to read the same scans again.
%Visualization order of presentation was counterbalanced.
They were asked to identify any stenosis they detected.
One participant was not able to finish the controlled experiment as the tutorial lasted longer than expected and they were paged to attend to a critical clinical case.
We have 6 answers for CV and none for 3D for this participant.
\REMOVE{This introduced an unbalanced number of repeated measures across participants which we addressed in our analysis.}

\subsection{Quantitative Analysis and Results}

\begin{figure}[! tb]
    \includegraphics[width=\linewidth]{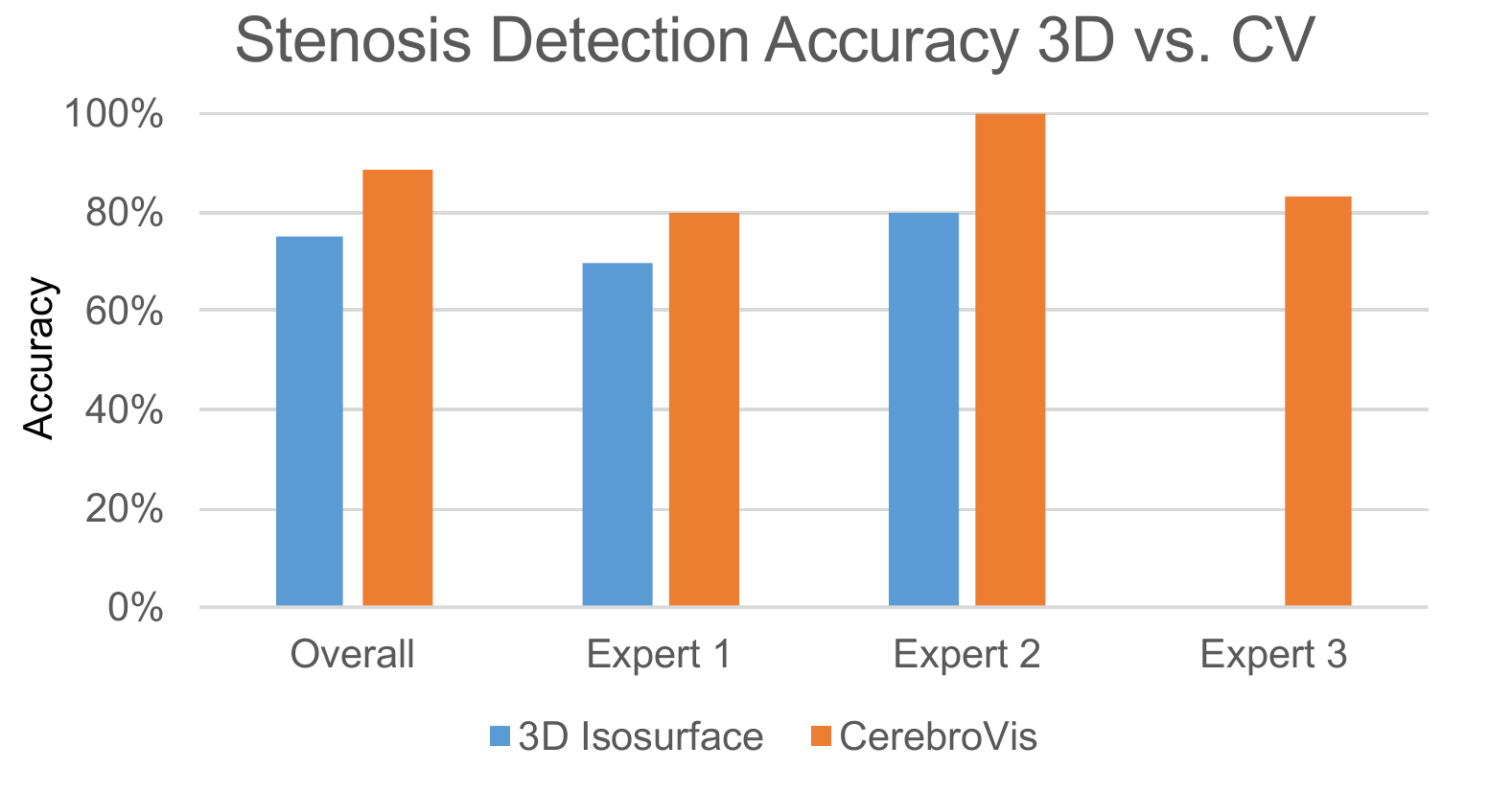}
    \caption{\REMOVE{Overview of participant data and the accuracy difference (total correct/total scans) between the 3D  visualization (3D) and 2D network layout CerebroVis (CV).
    One expert did not complete the 3D  visualization condition due to early departure from interview.}
    \ADD{Expert accuracy at identifying simulated stenoses using the 3D  visualization (3D) and 2D network layout CerebroVis (CV).
    One expert (\#3) did not complete the 3D  visualization condition due to early departure from the interview.}}
    \label{fig:result}
\end{figure}

We analyzed the $n=46$ (CV$=26$, 3D$=20$) answers from our three neuroradiologists using
\REMOVE{using a generalized estimating equation (GEE) logistic regression\cite{GEE}.
The GEE describes the marginal expectation of the discrete outcome variable---e.g., a binary true or false for whether a participant's answer is correct---as a function of the covariates.
Our covariate of interest is the visualization used: 3D (control) vs. CerebroVis (treatment).
Correlation among repeated within-subject observations is treated as a nuisance variable---i.e. another covariate, often referred to as the group.
The GEE makes no assumptions for the joint distribution of repeated measurements.
We chose an exchangeable working dependence for modeling within-subject correlations as we had unbalanced data and because it is more appropriate for short time intervals like our study\cite{RizoNotes}.
The focus of the GEE is on estimating the average response over the population and produces robust standard error estimates.
We used the GEE implementation in StatsModels\cite{seabold2010statsmodels}.}
\ADD{simple risk statistics, as we did not have enough participants to conduct a more rigorous statistical analysis.}
We compare the stenosis detection accuracy for the control 3D visualization (3D) and treatment CerebroVis (CV) in Fig.~\ref{fig:result}, overall and for individual experts.
Below we detail the results of our quantitative analysis including \REMOVE{odds ratio (OR) from the GEE; the 95\% confidence interval (CI); $p$-value from the Wald test of treatment significance; and for incorrect answers} the control absolute risk (ARC), treatment absolute risk (ART), and the absolute risk difference (ARD).
\ADD{Note that due to the small sample size these analyses should be interpreted as formative results.}
Our data and analysis code is available at \href{https://osf.io/e5sxt/}{osf.io/e5sxt}

\REMOVE{In general, we found that the neuroradiologists had better odds of correctly answering the diagnostic task questions using the 2D CerebroVis layout as compared to the 3D visualization \textit{(OR 2.5, 95\% CI 1.2--5.2, p 0.01, ARC 25.0\%, ART 11.5\%, ARD 13\%)} as well as correctly identifying an introduced stenosis (\textit{OR 2.3, 95\% CI 2.3--2.3, p 0.00, ARC 40.0\%, ART 7.1\%, ARD 32.8\%}).
Participants also had better odds of identifying false positives with CerebroVis, i.e., a feature the participant believes is a stenosis but that we did not introduce \textit{(OR 1.7, 95\% CI 1.1--2.6, p 0.01, ARC 10.0\%, ART 16.7\%, ARD -6.7\%)}.
Conversely, there was no observable difference between the odds for the 3D and CerebroVis conditions for false negatives, i.e., a participant did not correctly identify a stenosis that we did introduce \textit{(OR 0.2, 95\% CI 0.0--2.0, p 0.16, ARC 40.0\%, ART 7.1\%, ARD -32.9\%)}.
CerebroVis had worse odds than 3D for true negatives, i.e. the correct determination that a scan does not contain a stenosis \textit{(OR 0.8, 95\% CI 0.6--0.9, p 0.01, ARC 10.0\%, ART 16.7\% ARD -6.7\%)}.}
\REMOVE{\textbf{Summary:} Overall we found that identification of a confirmed stenosis is more accurate with CerebroVis (favourable true positive odds ratio).
But, with CerebroVis experts may classify subtle artery narrowing also as stenosis which contributes to an unfavorable true negative odds ratio and favorable false positive odds ratio.}
\ADD{In general, we found that the neuroradiologists were more likely to correctly answer the diagnostic task questions using the 2D CerebroVis layout compared to the 3D visualization \textit{(ARC 25\%, ART 12\%, ARD 13\%)} and more likely to correctly identify an introduced stenosis, i.e. a true positive (\textit{ARC 40\%, ART 7\%, ARD 33\%}).
However, participants were also more likely to identify false positives with CerebroVis, i.e., a feature the participant believes is a stenosis but that we did not introduce \textit{(ARC 10\%, ART 17\%, ARD -7\%)}.}
We reflect on these results in the discussion (Sec.~\ref{subsec:Validationdiscussion}).

\subsection{Qualitative Results}

We observed two consistent recurring themes in our interviews: 

\textbf{The linked view feature of the CerebroVis dashboard is essential to validate the identified abnormality with source data:} Participants were excited about the interaction between CerebroVis and the 2D isosurface projection of the brain arteries.
One of the participants said, \textit{``If I want I can spread out the arteries in 2D and then conveniently go back to the real data.''}
Another participant identified that the 2D isosurface projection was insufficient alone, but since the tool can link the abstract artery mark to the exact spatial position they could verify geometry in the source data.
They also said, \textit{``None of the existing visualization technique has the feature to link to the source data.''}

\textbf{The 2D layout has the potential to be an alternative to 3D representation:} All three participants saw potential in this novel cerebral artery visualization.
According to one, \textit{``This is already done by cardiologists to visualize stenosis, I am surprised we don't do it.''}
Another mentioned, \textit{``This is like a panoramic view of the arteries, which is an interesting engineering design.''}
When discussing cost-effectiveness one said, \textit{``2D visualization is probably going to be faster than the 3D visualization.''}

\section{Discussion}\label{discussion}

\subsection{Abstraction with Context}

\ADD{Through our design process we realized that the visualization of complex systems, like the cerebral artery network, can be intrinsically complicated due to the complex nature of the underlying data. In our iterative design process, we learned that the visualization of complex systems can be abstracted to represent information pertinent to domain goals and abstract tasks. For example, in a stenosis detection task, the information about artery width and the network topology outweighs the importance of features like artery length and exact spatial position of the arteries. To reduce the visual complexity, we designed a network representation which displayed information about the artery width and the topology of the cerebral branches (Fig.~\ref{designIteration} (3)) and de-emphasized non-required features such as artery length and the exact spatial position of arteries. While the experts appreciated the design, they did not understand how to interpret the non-spatially contextualized tree visualization back into 3D space. Ultimately, in this design study, we learned that abstract visualizations of complex systems could improve the task performance but there is a need to supplement the abstraction with user-centered context. During our design evolution, we found two visualization design paradigms that helped us embed context in  CerebroVis:}

\ADD{\textbf{1. Spatially Constrained Network Layout:} The cerebral artery network representation is spatially constrained to preserve the relative position of the arteries as per their location in the brain. The constraint provides a spatial context for the abstract topology visualization of the cerebral arteries. In our evaluation, we observed that the spatial context of cerebral arteries assisted the neuroradiologists in interpreting the abstract visualization. For example, in the stenosis detection task, experts correctly identified the site of the stenosis without the use of any legend. Thus user-centered spatial constraints embed the necessary context to understand the abstract representation of a complex artery network.}

\ADD{\textbf{2. Linked Views:}  Linked views can be used to explore multiple facets of a dataset~\cite{roberts2005exploratory}. In CerebroVis, the abstract visualization of the cerebral arteries does not preserve the exact spatial position and 3D geometry. Therefore, to preserve the 3D anatomical context of the data, the spatially constrained network visualization is linked with the 2D spatial projection of the brain data in our tool implementation. The link between the 3D spatial brain and abstract network visualization is essential for the clinical domain experts as it enables them to examine the geometry of the deformed artery in full context. For example, we do not encode the bends and curves of the arteries, but this information plays an important role in the treatment of vascular abnormalities. The availability of anatomical context through linked views allows an expert to validate the abnormality and thereby instills confidence in using the novel visual representation.}

\ADD{\textbf{Design Recommendations:}  In this design study we introduced a novel method to visualize the human cerebral artery network. While our visualization is tied explicitly to the human cerebral system, the design paradigms discussed in this study can be broadly applied outside of the brain to visualize other hierarchical circulatory systems. In this paper, we recommend the use of abstract topology visualization of the circulatory system to support  network tasks like path following and symmetry comparison. The familiarity of the abstract topology can be increased with relative spatial constraints to match the internal representation of the user mental model, and the abstract representation can be further linked with the spatially and anatomically accurate visual representation to allow in-depth analysis of vascular abnormalities.}

\vspace{-0.5 em}

\subsection{Evaluation Results}\label{subsec:Validationdiscussion}
Overall, we found that CerebroVis was easy to understand for all three study participants and they quickly adapted to the design.
The \REMOVE{comparative}\ADD{formative} quantitative evaluation results align with the qualitative feedback: overall CerebroVis was more effective for identifying stenoses in cerebral arteries.
Experts did a better job of identifying true positives (simulated stenosis cases) with the 2D network layout of CerebroVis as compared to a 3D isosurface visualization.
This validates our visualization against the primary design requirement that the visual encoding should support identification of abnormalities (Sec.~\ref{subsec:designrequirement}). 
In the evaluation we also notice that CerebroVis performs worse than the 3D representation for the detection of true negatives.
We speculate this is could be for two reasons.
First, the novel unfamiliar representation of CerebroVis, and the easy length comparisons it provides, may require additional training for experts to distinguish significant and non-significant differences in artery width.
Second, the presumably healthy patient scans we use may include undiagnosed stenoses that we did not introduce.

\section{\REMOVE{Future Work}\ADD{Limitations and Future Work}}
\REMOVE{
Neuroradiologists may need to visually communicate the results of diagnosis to patients, but with conventional 3D and 2D slice images, it can be a challenge to describe and explain the results to patients. Due to a simple design CerebroVis can be developed as an application to improve doctor-patient information exchange.Integrating CerebroVis with automated cerebrovascular abnormality detection systems can be beneficial for approximate tasks like symmetry comparison. Doctors use a rough visual comparison to detect differences, so there is a chance to miss out on some abnormalities. This can be easily avoided by providing annotated areas of symmetry variation, which can serve as cues for doctors to examine the highlighted area. CerebroVis can also be used to compare differences between network structure produced from different imaging techniques. MRA and CTA imaging techniques can give different granularity of information, and with an abstract 2D spatial layout it might be easier to spot these differences in the network structure. }

\ADD{
With CerebroVis we support symmetry comparison through the topology comparison of the cerebral arteries. The task of symmetry comparison in cerebral arteries can be further developed by layout design improvements or analytical integration to include more spatial information. For example, the layout could preserve the exact spatial distance between two arteries, branching sites could preserve the angle of branching, or the network edges could encode tortuosity of the arteries. Spatial information will benefit the overall visual comparison, but the effect on the other tasks such as stenosis detection also needs to be carefully considered. Integrating CerebroVis with automated cerebrovascular abnormality detection systems may be beneficial for approximate tasks including symmetry comparison. Doctors use a rough visual comparison to detect differences, thus an abnormality may elude detection. This could be avoided by providing annotated areas of symmetry variation to serve as cues for doctors to examine the highlighted area.  Finally, in support of sometimes small artery-width abnormalities, such as aneurysms, instead of encoding edge thickness proportional to the average width of the entire artery the visual encoding could display a variable-width encoding where needed to highlight such outliers.}%Currently, CerebroVis encodes edge thickness propotional to the average width of the entire artery. However, an artery may span multiple segments and thus it is possible to miss out non-significant deviations. Future work can encode the artery width encoding in more detail, by ...................?  }

\section{Conclusion}
We present CerebroVis, a novel abstract but spatially contextualized network layout for visualizing cerebral artery networks.
We also contribute a novel framing and network theory definition of the cerebral artery system.
Through expert interview and observations we characterize the domain goals in an ordered list of importance and present them as abstract visualization and network analysis tasks.
We evaluate the layout and the co-developed visualization prototype through a mixed methods study with three neuroradiologists.
In a controlled task-based study we found that our abstract visualization improved task performance over a conventional 3D visualization for identifying intracranial artery stenosis.
From semi-structured interviews we determined that the inclusion of spatial context helped preserve the users' mental maps of the underlying geometry.
\REMOVE{More broadly, we believe visualization design for other hierarchical circulatory system components outside the brain---as well as researchers in other domains with 3D data---could benefit from our design approach of abstraction with context.}
\ADD{More broadly, we believe visualization design for other hierarchical circulatory system components outside the brain could benefit from our design methodology.}

\acknowledgments{
We thank David Saffo, Laura South, Michail Schwab, Sara Di Bartolomeo, and Yixuan ``Janice'' Zhang for their valuable feedback. A special thanks to Ian Helgi Magnusson for editing and proof reading help. This work was supported by NSF CISE CRII award no. 1657466.}

\bibliographystyle{abbrv-doi-hyperref}

\bibliography{template}

\begin{thebibliography}{10}

\bibitem{Alper:2013:WGC:2470654.2470724}
\href{https://doi.org/10.1145/2470654.2470724}{B.~Alper, B.~Bach,
  N.~Henry~Riche, T.~Isenberg, and J.-D. Fekete}.
\newblock \href{https://doi.org/10.1145/2470654.2470724}{Weighted graph
  comparison techniques for brain connectivity analysis}.
\newblock \href{https://doi.org/10.1145/2470654.2470724}{In {\em Proceedings of
  the SIGCHI Conference on Human Factors in Computing Systems}},
  \href{https://doi.org/10.1145/2470654.2470724}{CHI '13},
  \href{https://doi.org/10.1145/2470654.2470724}{pp. 483--492}.
  \href{https://doi.org/10.1145/2470654.2470724}{ACM},
  \href{https://doi.org/10.1145/2470654.2470724}{New York, NY, USA},
  \href{https://doi.org/10.1145/2470654.2470724}{2013}.
  \href{https://doi.org/10.1145/2470654.2470724}
{doi: {{%
10\hspace{.1pt}\discretionary{.}{%
}{.}\hspace{.4pt}1145\discretionary{/}{%
}{/}2470654\hspace{.1pt}\discretionary{.}{%
}{.}\hspace{.4pt}2470724}}}


\bibitem{Aviv1975}
\href{https://doi.org/10.3174/ajnr.A0689}{R.~Aviv, J.~Mandelcorn,
  S.~Chakraborty, D.~Gladstone, S.~Malham, G.~Tomlinson, A.~Fox, and
  S.~Symons}.
\newblock \href{https://doi.org/10.3174/ajnr.A0689}{Alberta stroke program
  early ct scoring of ct perfusion in early stroke visualization and
  assessment}.
\newblock \href{https://doi.org/10.3174/ajnr.A0689}{{\em American Journal of
  Neuroradiology}},
  \href{https://doi.org/10.3174/ajnr.A0689}{28(10):1975--1980},
  \href{https://doi.org/10.3174/ajnr.A0689}{nov 2007}.
  \href{https://doi.org/10.3174/ajnr.A0689}
{doi: {{%
10\hspace{.1pt}\discretionary{.}{%
}{.}\hspace{.4pt}3174\discretionary{/}{%
}{/}ajnr\hspace{.1pt}\discretionary{.}{%
}{.}\hspace{.4pt}A0689}}}


\bibitem{Aydin2011}
\href{https://doi.org/10.1214/11-ejs612}{B.~Ayd{\i}n, G.~Pataki, H.~Wang,
  A.~Ladha, E.~Bullitt, and J.~Marron}.
\newblock \href{https://doi.org/10.1214/11-ejs612}{Visualizing the structure of
  large trees}.
\newblock \href{https://doi.org/10.1214/11-ejs612}{{\em Electronic Journal of
  Statistics}}, \href{https://doi.org/10.1214/11-ejs612}{5(0):405--420},
  \href{https://doi.org/10.1214/11-ejs612}{2011}.
  \href{https://doi.org/10.1214/11-ejs612}
{doi: {{%
10\hspace{.1pt}\discretionary{.}{%
}{.}\hspace{.4pt}1214\discretionary{/}{%
}{/}11\discretionary{%
}{-}{-}ejs612}}}


\bibitem{barsky2008cerebral}
A.~Barsky, T.~Munzner, J.~Gardy, and R.~Kincaid.
\newblock Cerebral: Visualizing multiple experimental conditions on a graph
  with biological context.
\newblock {\em IEEE transactions on visualization and computer graphics},
  14(6):1253--1260, 2008.

\bibitem{battista1998graph}
G.~D. Battista, P.~Eades, R.~Tamassia, and I.~G. Tollis.
\newblock {\em Graph drawing: algorithms for the visualization of graphs}.
\newblock Prentice Hall PTR, 1998.

\bibitem{6065015}
\href{https://doi.org/10.1109/TVCG.2011.192}{M.~Borkin, K.~Gajos, A.~Peters,
  D.~Mitsouras, S.~Melchionna, F.~Rybicki, C.~Feldman, and H.~Pfister}.
\newblock \href{https://doi.org/10.1109/TVCG.2011.192}{Evaluation of artery
  visualizations for heart disease diagnosis}.
\newblock \href{https://doi.org/10.1109/TVCG.2011.192}{{\em IEEE Transactions
  on Visualization and Computer Graphics}},
  \href{https://doi.org/10.1109/TVCG.2011.192}{17(12):2479--2488},
  \href{https://doi.org/10.1109/TVCG.2011.192}{Dec 2011}.
  \href{https://doi.org/10.1109/TVCG.2011.192}
{doi: {{%
10\hspace{.1pt}\discretionary{.}{%
}{.}\hspace{.4pt}1109\discretionary{/}{%
}{/}TVCG\hspace{.1pt}\discretionary{.}{%
}{.}\hspace{.4pt}2011\hspace{.1pt}\discretionary{.}{%
}{.}\hspace{.4pt}192}}}


\bibitem{Bostock:2011:DDD:2068462.2068631}
\href{https://doi.org/10.1109/TVCG.2011.185}{M.~Bostock, V.~Ogievetsky, and
  J.~Heer}.
\newblock \href{https://doi.org/10.1109/TVCG.2011.185}{D3 data-driven
  documents}.
\newblock \href{https://doi.org/10.1109/TVCG.2011.185}{{\em IEEE Transactions
  on Visualization and Computer Graphics}},
  \href{https://doi.org/10.1109/TVCG.2011.185}{17(12):2301--2309},
  \href{https://doi.org/10.1109/TVCG.2011.185}{Dec. 2011}.
  \href{https://doi.org/10.1109/TVCG.2011.185}
{doi: {{%
10\hspace{.1pt}\discretionary{.}{%
}{.}\hspace{.4pt}1109\discretionary{/}{%
}{/}TVCG\hspace{.1pt}\discretionary{.}{%
}{.}\hspace{.4pt}2011\hspace{.1pt}\discretionary{.}{%
}{.}\hspace{.4pt}185}}}


\bibitem{Brehmer2013}
M.~Brehmer and T.~Munzner.
\newblock {A multi-level typology of abstract visualization tasks}.
\newblock {\em IEEE Trans. Visualization and Computer Graphics (TVCG) (Proc.
  InfoVis)}, 19(12):2376--2385, 2013.

\bibitem{burch2011evaluation}
M.~Burch, N.~Konevtsova, J.~Heinrich, M.~Hoeferlin, and D.~Weiskopf.
\newblock Evaluation of traditional, orthogonal, and radial tree diagrams by an
  eye tracking study.
\newblock {\em IEEE Transactions on Visualization and Computer Graphics},
  17(12):2440--2448, 2011.

\bibitem{ischemicstroke}
{Centers for Disease Control and Prevention}.
\newblock Cdc stroke information.
\newblock \url{https://www.cdc.gov/stroke/types_of_stroke.html}.

\bibitem{Coleman96}
\href{https://doi.org/10.1002/(SICI)1097-024X(199612)26:12<1415::AID-SPE69>3.3.CO;2-G}{M.~K.
  Coleman and D.~S. Parker}.
\newblock
  \href{https://doi.org/10.1002/(SICI)1097-024X(199612)26:12<1415::AID-SPE69>3.3.CO;2-G}{Aesthetics-based
  graph layout for human consumption}.
\newblock
  \href{https://doi.org/10.1002/(SICI)1097-024X(199612)26:12<1415::AID-SPE69>3.3.CO;2-G}{{\em
  Softw. Pract. Exper.}},
  \href{https://doi.org/10.1002/(SICI)1097-024X(199612)26:12<1415::AID-SPE69>3.3.CO;2-G}{26(12):1415--1438},
  \href{https://doi.org/10.1002/(SICI)1097-024X(199612)26:12<1415::AID-SPE69>3.3.CO;2-G}{Dec.
  1996}.
  \href{https://doi.org/10.1002/(SICI)1097-024X(199612)26:12<1415::AID-SPE69>3.3.CO;2-G}
{doi: {{%
10\hspace{.1pt}\discretionary{.}{%
}{.}\hspace{.4pt}1002\discretionary{/}{%
}{/}\discretionary{%
}{(}{(}SICI\discretionary{)}{%
}{)}1097\discretionary{%
}{-}{-}024X\discretionary{%
}{(}{(}199612\discretionary{)}{%
}{)}26\discretionary{:}{%
}{:}12{\textless}1415\discretionary{:}{%
}{:}\discretionary{:}{%
}{:}AID\discretionary{%
}{-}{-}SPE69{\textgreater}3\hspace{.1pt}\discretionary{.}{%
}{.}\hspace{.4pt}3\hspace{.1pt}\discretionary{.}{%
}{.}\hspace{.4pt}CO\discretionary{;}{%
}{;}2\discretionary{%
}{-}{-}G}}}


\bibitem{csermely2013structure}
P.~Csermely, T.~Korcsm{\'a}ros, H.~J. Kiss, G.~London, and R.~Nussinov.
\newblock Structure and dynamics of molecular networks: a novel paradigm of
  drug discovery: a comprehensive review.
\newblock {\em Pharmacology \& therapeutics}, 138(3):333--408, 2013.

\bibitem{davidson1996drawing}
R.~Davidson and D.~Harel.
\newblock Drawing graphs nicely using simulated annealing.
\newblock {\em ACM Transactions on Graphics (TOG)}, 15(4):301--331, 1996.

\bibitem{DelhiMetro}
{Delhi Metro Rail Corporation Ltd.}
\newblock \url{http://www.delhimetrorail.com/Zoom_Map.aspx}.

\bibitem{despotovic2015mri}
I.~Despotovi{\'c}, B.~Goossens, and W.~Philips.
\newblock Mri segmentation of the human brain: challenges, methods, and
  applications.
\newblock {\em Computational and mathematical methods in medicine}, 2015, 2015.

\bibitem{Nmap}
\href{https://doi.org/10.1109/TVCG.2014.2346276}{F.~S. L.~G. {Duarte},
  F.~{Sikansi}, F.~M. {Fatore}, S.~G. {Fadel}, and F.~V. {Paulovich}}.
\newblock \href{https://doi.org/10.1109/TVCG.2014.2346276}{Nmap: A novel
  neighborhood preservation space-filling algorithm}.
\newblock \href{https://doi.org/10.1109/TVCG.2014.2346276}{{\em IEEE
  Transactions on Visualization and Computer Graphics}},
  \href{https://doi.org/10.1109/TVCG.2014.2346276}{20(12):2063--2071},
  \href{https://doi.org/10.1109/TVCG.2014.2346276}{Dec 2014}.
  \href{https://doi.org/10.1109/TVCG.2014.2346276}
{doi: {{%
10\hspace{.1pt}\discretionary{.}{%
}{.}\hspace{.4pt}1109\discretionary{/}{%
}{/}TVCG\hspace{.1pt}\discretionary{.}{%
}{.}\hspace{.4pt}2014\hspace{.1pt}\discretionary{.}{%
}{.}\hspace{.4pt}2346276}}}


\bibitem{Dunne15Readabilitymetricfeedback}
\href{https://doi.org/10.1147/JRD.2015.2411412}{C.~Dunne, S.~I. Ross,
  B.~Shneiderman, and M.~Martino}.
\newblock \href{https://doi.org/10.1147/JRD.2015.2411412}{{R}eadability metric
  feedback for aiding node-link visualization designers}.
\newblock \href{https://doi.org/10.1147/JRD.2015.2411412}{{\em {IBM} {J}ournal
  of {R}esearch and {D}evelopment}},
  \href{https://doi.org/10.1147/JRD.2015.2411412}{59(2/3):14:1--14:16},
  \href{https://doi.org/10.1147/JRD.2015.2411412}{Mar. 2015}.
  \href{https://doi.org/10.1147/JRD.2015.2411412}
{doi: {{%
10\hspace{.1pt}\discretionary{.}{%
}{.}\hspace{.4pt}1147\discretionary{/}{%
}{/}JRD\hspace{.1pt}\discretionary{.}{%
}{.}\hspace{.4pt}2015\hspace{.1pt}\discretionary{.}{%
}{.}\hspace{.4pt}2411412}}}


\bibitem{dykes1998cartographic}
J.~Dykes.
\newblock Cartographic visualization.
\newblock {\em Journal of the Royal Statistical Society: Series D (The
  Statistician)}, 47(3):485--497, 1998.

\bibitem{eades1989draw}
P.~Eades and L.~Xuemin.
\newblock How to draw a directed graph.
\newblock In {\em Visual Languages, 1989., IEEE Workshop on}, pp. 13--17. IEEE,
  1989.

\bibitem{ferrari1969drawing}
D.~Ferrari and L.~Mezzalira.
\newblock {\em On drawing a graph with the minimum number of crossings}.
\newblock Politecnico, 1969.

\bibitem{fishman2006volume}
E.~K. Fishman, D.~R. Ney, D.~G. Heath, F.~M. Corl, K.~M. Horton, and P.~T.
  Johnson.
\newblock Volume rendering versus maximum intensity projection in ct
  angiography: what works best, when, and why.
\newblock {\em Radiographics}, 26(3):905--922, 2006.

\bibitem{fruchterman1991graph}
T.~M. Fruchterman and E.~M. Reingold.
\newblock Graph drawing by force-directed placement.
\newblock {\em Software: Practice and experience}, 21(11):1129--1164, 1991.

\bibitem{gacs1983ct}
G.~Gacs, A.~Fox, H.~Barnett, and F.~Vinuela.
\newblock Ct visualization of intracranial arterial thromboembolism.
\newblock {\em Stroke}, 14(5):756--762, 1983.

\bibitem{ghoniem2015weighted}
M.~Ghoniem, M.~Cornil, B.~Broeksema, M.~Stefas, and B.~Otjacques.
\newblock Weighted maps: treemap visualization of geolocated quantitative data.
\newblock In {\em Visualization and Data Analysis 2015}, vol. 9397, p. 93970G.
  International Society for Optics and Photonics, 2015.

\bibitem{goguen1993techniques}
J.~A. Goguen and C.~Linde.
\newblock Techniques for requirements elicitation.
\newblock In {\em [1993] Proceedings of the IEEE International Symposium on
  Requirements Engineering}, pp. 152--164. IEEE, 1993.

\bibitem{grabler2008automatic}
F.~Grabler, M.~Agrawala, R.~W. Sumner, and M.~Pauly.
\newblock {\em Automatic generation of tourist maps}, vol.~27.
\newblock ACM, 2008.

\bibitem{guerra2013visualizing}
J.~Guerra-G{\'o}mez, M.~L. Pack, C.~Plaisant, and B.~Shneiderman.
\newblock Visualizing change over time using dynamic hierarchies: Treeversity2
  and the stemview.
\newblock {\em IEEE Transactions on Visualization and Computer Graphics},
  19(12):2566--2575, 2013.

\bibitem{guerra2013treeversity}
J.~A. Guerra-G{\'o}mez, A.~Buck-Coleman, M.~L. Pack, C.~Plaisant, and
  B.~Shneiderman.
\newblock Treeversity: interactive visualizations for comparing hierarchical
  data sets.
\newblock {\em Transportation Research Record}, 2392(1):48--58, 2013.

\bibitem{gutwenger2003new}
C.~Gutwenger, M.~J{\"u}nger, K.~Klein, J.~Kupke, S.~Leipert, and P.~Mutzel.
\newblock A new approach for visualizing uml class diagrams.
\newblock In {\em Proceedings of the 2003 ACM symposium on Software
  visualization}, pp. 179--188. ACM, 2003.

\bibitem{he1998constrained}
W.~He and K.~Marriott.
\newblock Constrained graph layout.
\newblock {\em Constraints}, 3(4):289--314, 1998.

\bibitem{henry2007nodetrix}
N.~Henry, J.-D. Fekete, and M.~J. McGuffin.
\newblock Nodetrix: a hybrid visualization of social networks.
\newblock {\em IEEE transactions on visualization and computer graphics},
  13(6):1302--1309, 2007.

\bibitem{SWCv}
{HHMIs Janelia Research Campus}.
\newblock {SharkViewer}.
\newblock \url{https://www.janelia.org/sharkviewer}.

\bibitem{holten2008visual}
D.~Holten and J.~J. Van~Wijk.
\newblock Visual comparison of hierarchically organized data.
\newblock In {\em Computer Graphics Forum}, vol.~27, pp. 759--766. Wiley Online
  Library, 2008.

\bibitem{johnson2004mixed}
R.~B. Johnson and A.~J. Onwuegbuzie.
\newblock Mixed methods research: A research paradigm whose time has come.
\newblock {\em Educational researcher}, 33(7):14--26, 2004.

\bibitem{johnson2016stroke}
W.~Johnson, O.~Onuma, M.~Owolabi, and S.~Sachdev.
\newblock Stroke: a global response is needed.
\newblock {\em Bulletin of the World Health Organization}, 94(9):634, 2016.

\bibitem{neurolines}
A.~K.~Al-Awami, J.~Beyer, H.~Strobelt, N.~Kasthuri, J.~Lichtman, H.~Pfister,
  and M.~Hadwiger.
\newblock Neurolines: A subway map metaphor for visualizing nanoscale neuronal
  connectivity.
\newblock 20:2369--2378, 12 2014.

\bibitem{Kobourov:2015:GPG:2951136.2951162}
\href{https://doi.org/10.1007/978-3-319-27261-0_50}{S.~G. Kobourov,
  T.~Mchedlidze, and L.~Vonessen}.
\newblock \href{https://doi.org/10.1007/978-3-319-27261-0_50}{Gestalt
  principles in graph drawing}.
\newblock \href{https://doi.org/10.1007/978-3-319-27261-0_50}{In {\em Revised
  Selected Papers of the 23rd International Symposium on Graph Drawing and
  Network Visualization - Volume 9411}},
  \href{https://doi.org/10.1007/978-3-319-27261-0_50}{GD 2015},
  \href{https://doi.org/10.1007/978-3-319-27261-0_50}{pp. 558--560}.
  \href{https://doi.org/10.1007/978-3-319-27261-0_50}{Springer-Verlag New York,
  Inc.}, \href{https://doi.org/10.1007/978-3-319-27261-0_50}{New York, NY,
  USA}, \href{https://doi.org/10.1007/978-3-319-27261-0_50}{2015}.
  \href{https://doi.org/10.1007/978-3-319-27261-0_50}
{doi: {{%
10\hspace{.1pt}\discretionary{.}{%
}{.}\hspace{.4pt}1007\discretionary{/}{%
}{/}978\discretionary{%
}{-}{-}3\discretionary{%
}{-}{-}319\discretionary{%
}{-}{-}27261\discretionary{%
}{-}{-}0\_50}}}


\bibitem{krzywinski2011hive}
M.~Krzywinski, I.~Birol, S.~J. Jones, and M.~A. Marra.
\newblock Hive plots rational approach to visualizing networks.
\newblock {\em Briefings in bioinformatics}, 13(5):627--644, 2011.

\bibitem{kubisch2012vessel}
C.~Kubisch, S.~Gla{\ss}er, M.~Neugebauer, and B.~Preim.
\newblock Vessel visualization with volume rendering.
\newblock In {\em Visualization in Medicine and Life Sciences II}, pp.
  109--132. Springer, 2012.

\bibitem{lee2006task}
B.~Lee, C.~Plaisant, C.~S. Parr, J.-D. Fekete, and N.~Henry.
\newblock Task taxonomy for graph visualization.
\newblock In {\em Proceedings of the 2006 AVI workshop on BEyond time and
  errors: novel evaluation methods for information visualization}, pp. 1--5.
  ACM, 2006.

\bibitem{Lev2736}
\href{https://doi.org/10.1161/01.STR.0000041999.64363.B2}{M.~H. Lev, W.~J.
  Koroshetz, L.~H. Schwamm, and R.~G. Gonzalez}.
\newblock \href{https://doi.org/10.1161/01.STR.0000041999.64363.B2}{Ct or mri
  for imaging patients with acute stroke: Visualization of
  {\textquotedblleft}tissue at risk{\textquotedblright}?}
\newblock \href{https://doi.org/10.1161/01.STR.0000041999.64363.B2}{{\em
  Stroke}},
  \href{https://doi.org/10.1161/01.STR.0000041999.64363.B2}{33(12):2736--2737},
  \href{https://doi.org/10.1161/01.STR.0000041999.64363.B2}{2002}.
  \href{https://doi.org/10.1161/01.STR.0000041999.64363.B2}
{doi: {{%
10\hspace{.1pt}\discretionary{.}{%
}{.}\hspace{.4pt}1161\discretionary{/}{%
}{/}01\hspace{.1pt}\discretionary{.}{%
}{.}\hspace{.4pt}STR\hspace{.1pt}\discretionary{.}{%
}{.}\hspace{.4pt}0000041999\hspace{.1pt}\discretionary{.}{%
}{.}\hspace{.4pt}64363\hspace{.1pt}\discretionary{.}{%
}{.}\hspace{.4pt}B2}}}


\bibitem{10.1093/neuros/nyx325}
\href{https://doi.org/10.1093/neuros/nyx325}{A.~Lin, R.~Agid, S.~Rawal, and
  D.~M. Mandell}.
\newblock \href{https://doi.org/10.1093/neuros/nyx325}{{Cerebrovascular
  Imaging: Which Test is Best?}}
\newblock \href{https://doi.org/10.1093/neuros/nyx325}{{\em Neurosurgery}},
  \href{https://doi.org/10.1093/neuros/nyx325}{83(1):5--18},
  \href{https://doi.org/10.1093/neuros/nyx325}{06 2017}.
  \href{https://doi.org/10.1093/neuros/nyx325}
{doi: {{%
10\hspace{.1pt}\discretionary{.}{%
}{.}\hspace{.4pt}1093\discretionary{/}{%
}{/}neuros\discretionary{/}{%
}{/}nyx325}}}


\bibitem{lipton1985method}
R.~J. Lipton, S.~C. North, and J.~S. Sandberg.
\newblock A method for drawing graphs.
\newblock In {\em Proceedings of the first annual symposium on Computational
  geometry}, pp. 153--160. ACM, 1985.

\bibitem{BostonSubway}
{Massachussets Bay Transport Authority}.
\newblock \url{https://www.mbta.com/schedules/subway}.

\bibitem{Mohammed2017}
H.~Mohammed, A.~K. Al-Awami, J.~Beyer, C.~Cali, P.~Magistretti, H.~Pfister, and
  M.~Hadwiger.
\newblock Abstractocyte: A visual tool for exploring nanoscale astroglial
  cells.
\newblock 2017.

\bibitem{5290695}
\href{https://doi.org/10.1109/TVCG.2009.111}{T.~{Munzner}}.
\newblock \href{https://doi.org/10.1109/TVCG.2009.111}{A nested model for
  visualization design and validation}.
\newblock \href{https://doi.org/10.1109/TVCG.2009.111}{{\em IEEE Transactions
  on Visualization and Computer Graphics}},
  \href{https://doi.org/10.1109/TVCG.2009.111}{15(6):921--928},
  \href{https://doi.org/10.1109/TVCG.2009.111}{Nov 2009}.
  \href{https://doi.org/10.1109/TVCG.2009.111}
{doi: {{%
10\hspace{.1pt}\discretionary{.}{%
}{.}\hspace{.4pt}1109\discretionary{/}{%
}{/}TVCG\hspace{.1pt}\discretionary{.}{%
}{.}\hspace{.4pt}2009\hspace{.1pt}\discretionary{.}{%
}{.}\hspace{.4pt}111}}}


\bibitem{munzner2003treejuxtaposer}
T.~Munzner, F.~Guimbreti{\`e}re, S.~Tasiran, L.~Zhang, and Y.~Zhou.
\newblock Treejuxtaposer: scalable tree comparison using focus+ context with
  guaranteed visibility.
\newblock In {\em ACM Transactions on Graphics (TOG)}, vol.~22, pp. 453--462.
  ACM, 2003.

\bibitem{purchase1997aesthetic}
H.~Purchase.
\newblock Which aesthetic has the greatest effect on human understanding?
\newblock In {\em International Symposium on Graph Drawing}, pp. 248--261.
  Springer, 1997.

\bibitem{purchase2006important}
H.~C. Purchase, E.~Hoggan, and C.~G{\"o}rg.
\newblock How important is the ``mental map"?--an empirical investigation of a
  dynamic graph layout algorithm.
\newblock In {\em International Symposium on Graph Drawing}, pp. 184--195.
  Springer, 2006.

\bibitem{reingold1981tidier}
E.~M. Reingold and J.~S. Tilford.
\newblock Tidier drawings of trees.
\newblock {\em IEEE Transactions on software Engineering}, (2):223--228, 1981.

\bibitem{roberts2005exploratory}
J.~C. Roberts.
\newblock Exploratory visualization with multiple linked views.
\newblock In {\em Exploring geovisualization}, pp. 159--180. Elsevier, 2005.

\bibitem{rothwell2012junior}
C.~Rothwell, B.~Burford, J.~Morrison, G.~Morrow, M.~Allen, C.~Davies,
  B.~Baldauf, J.~Spencer, N.~Johnson, E.~Peile, et~al.
\newblock Junior doctors prescribing: enhancing their learning in practice.
\newblock {\em British journal of clinical pharmacology}, 73(2):194--202, 2012.

\bibitem{schreiber2009generic}
F.~Schreiber, T.~Dwyer, K.~Marriott, and M.~Wybrow.
\newblock A generic algorithm for layout of biological networks.
\newblock {\em BMC bioinformatics}, 10(1):375, 2009.

\bibitem{5473227}
\href{https://doi.org/10.1109/TVCG.2010.79}{H.~{Schulz}, S.~{Hadlak}, and
  H.~{Schumann}}.
\newblock \href{https://doi.org/10.1109/TVCG.2010.79}{The design space of
  implicit hierarchy visualization: A survey}.
\newblock \href{https://doi.org/10.1109/TVCG.2010.79}{{\em IEEE Transactions on
  Visualization and Computer Graphics}},
  \href{https://doi.org/10.1109/TVCG.2010.79}{17(4):393--411},
  \href{https://doi.org/10.1109/TVCG.2010.79}{April 2011}.
  \href{https://doi.org/10.1109/TVCG.2010.79}
{doi: {{%
10\hspace{.1pt}\discretionary{.}{%
}{.}\hspace{.4pt}1109\discretionary{/}{%
}{/}TVCG\hspace{.1pt}\discretionary{.}{%
}{.}\hspace{.4pt}2010\hspace{.1pt}\discretionary{.}{%
}{.}\hspace{.4pt}79}}}


\bibitem{schulz2011treevis}
H.-J. Schulz.
\newblock Treevis. net: A tree visualization reference.
\newblock {\em IEEE Computer Graphics and Applications}, 31(6):11--15, 2011.

\bibitem{sedlmair2012design}
M.~Sedlmair, M.~Meyer, and T.~Munzner.
\newblock Design study methodology: Reflections from the trenches and the
  stacks.
\newblock {\em IEEE transactions on visualization and computer graphics},
  18(12):2431--2440, 2012.

\bibitem{shneiderman1998treemaps}
B.~Shneiderman and C.~Plaisant.
\newblock Treemaps for space-constrained visualization of hierarchies.
\newblock 1998.

\bibitem{skwerer2014tree}
S.~Skwerer, E.~Bullitt, S.~Huckemann, E.~Miller, I.~Oguz, M.~Owen,
  V.~Patrangenaru, S.~Provan, and J.~Marron.
\newblock Tree-oriented analysis of brain artery structure.
\newblock {\em Journal of mathematical imaging and vision}, 50(1-2):126--143,
  2014.

\bibitem{Sugiyama02GraphDrawingand}
\href{http://books.google.com/books?id=H06j5GJKITIC}{K.~Sugiyama}.
\newblock \href{http://books.google.com/books?id=H06j5GJKITIC}{{\em {G}raph
  drawing and applications for software and knowledge engineers}},
  \href{http://books.google.com/books?id=H06j5GJKITIC}{vol.~11 of {\em Series
  on Software Engineering and Knowledge Engineering}}.
\newblock \href{http://books.google.com/books?id=H06j5GJKITIC}{World Scientific
  Publishing Company},
  \href{http://books.google.com/books?id=H06j5GJKITIC}{June 2002}.

\bibitem{Sugiyama81}
\href{https://doi.org/10.1109/TSMC.1981.4308636}{K.~Sugiyama, S.~Tagawa, and
  M.~Toda}.
\newblock \href{https://doi.org/10.1109/TSMC.1981.4308636}{Methods for visual
  understanding of hierarchical system structures}.
\newblock \href{https://doi.org/10.1109/TSMC.1981.4308636}{{\em IEEE
  Transactions on Systems, Man, and Cybernetics}},
  \href{https://doi.org/10.1109/TSMC.1981.4308636}{11(2):109--125},
  \href{https://doi.org/10.1109/TSMC.1981.4308636}{Feb 1981}.
  \href{https://doi.org/10.1109/TSMC.1981.4308636}
{doi: {{%
10\hspace{.1pt}\discretionary{.}{%
}{.}\hspace{.4pt}1109\discretionary{/}{%
}{/}TSMC\hspace{.1pt}\discretionary{.}{%
}{.}\hspace{.4pt}1981\hspace{.1pt}\discretionary{.}{%
}{.}\hspace{.4pt}4308636}}}


\bibitem{tanaka2008chronic}
M.~Tanaka, E.~Shimosegawa, K.~Kajimoto, Y.~Kimura, H.~Kato, N.~Oku, M.~Hori,
  K.~Kitagawa, and J.~Hatazawa.
\newblock Chronic middle cerebral artery occlusion: a hemodynamic and metabolic
  study with positron-emission tomography.
\newblock {\em American Journal of Neuroradiology}, 29(10):1841--1846, 2008.

\bibitem{Colajs}
{Tim Dwyer}.
\newblock cola.js:constraint-based layout in the browser.
\newblock \url{https://ialab.it.monash.edu/webcola/}.

\bibitem{MIP}
\href{https://doi.org/10.1109/42.41482}{J.~W. Wallis, T.~R. Miller, C.~A.
  Lerner, and E.~C. Kleerup}.
\newblock \href{https://doi.org/10.1109/42.41482}{Three-dimensional display in
  nuclear medicine}.
\newblock \href{https://doi.org/10.1109/42.41482}{{\em IEEE Transactions on
  Medical Imaging}}, \href{https://doi.org/10.1109/42.41482}{8(4):297--230},
  \href{https://doi.org/10.1109/42.41482}{Dec 1989}.
  \href{https://doi.org/10.1109/42.41482}
{doi: {{%
10\hspace{.1pt}\discretionary{.}{%
}{.}\hspace{.4pt}1109\discretionary{/}{%
}{/}42\hspace{.1pt}\discretionary{.}{%
}{.}\hspace{.4pt}41482}}}


\bibitem{ware2002cognitive}
\href{https://doi.org/10.1057\%2Fpalgrave.ivs.9500013}{C.~Ware, H.~Purchase,
  L.~Colpoys, and M.~McGill}.
\newblock \href{https://doi.org/10.1057\%2Fpalgrave.ivs.9500013}{Cognitive
  measurements of graph aesthetics}.
\newblock \href{https://doi.org/10.1057\%2Fpalgrave.ivs.9500013}{{\em
  Information visualization}},
  \href{https://doi.org/10.1057\%2Fpalgrave.ivs.9500013}{1(2):103--110},
  \href{https://doi.org/10.1057\%2Fpalgrave.ivs.9500013}{2002}.
  \href{https://doi.org/10.1057\%2Fpalgrave.ivs.9500013}
{doi: {{%
10\hspace{.1pt}\discretionary{.}{%
}{.}\hspace{.4pt}1057\%2Fpalgrave\hspace{.1pt}\discretionary{.}{%
}{.}\hspace{.4pt}ivs\hspace{.1pt}\discretionary{.}{%
}{.}\hspace{.4pt}9500013}}}


\bibitem{SOT}
\href{https://doi.org/10.1109/TVCG.2008.165}{J.~{Wood} and J.~{Dykes}}.
\newblock \href{https://doi.org/10.1109/TVCG.2008.165}{Spatially ordered
  treemaps}.
\newblock \href{https://doi.org/10.1109/TVCG.2008.165}{{\em IEEE Transactions
  on Visualization and Computer Graphics}},
  \href{https://doi.org/10.1109/TVCG.2008.165}{14(6):1348--1355},
  \href{https://doi.org/10.1109/TVCG.2008.165}{Nov 2008}.
  \href{https://doi.org/10.1109/TVCG.2008.165}
{doi: {{%
10\hspace{.1pt}\discretionary{.}{%
}{.}\hspace{.4pt}1109\discretionary{/}{%
}{/}TVCG\hspace{.1pt}\discretionary{.}{%
}{.}\hspace{.4pt}2008\hspace{.1pt}\discretionary{.}{%
}{.}\hspace{.4pt}165}}}


\bibitem{Worsley913}
\href{https://doi.org/10.1098/rstb.2005.1637}{K.~J. Worsley, J.-I. Chen,
  J.~Lerch, and A.~C. Evans}.
\newblock \href{https://doi.org/10.1098/rstb.2005.1637}{Comparing functional
  connectivity via thresholding correlations and singular value decomposition}.
\newblock \href{https://doi.org/10.1098/rstb.2005.1637}{{\em Philosophical
  Transactions of the Royal Society of London B: Biological Sciences}},
  \href{https://doi.org/10.1098/rstb.2005.1637}{360(1457):913--920},
  \href{https://doi.org/10.1098/rstb.2005.1637}{2005}.
  \href{https://doi.org/10.1098/rstb.2005.1637}
{doi: {{%
10\hspace{.1pt}\discretionary{.}{%
}{.}\hspace{.4pt}1098\discretionary{/}{%
}{/}rstb\hspace{.1pt}\discretionary{.}{%
}{.}\hspace{.4pt}2005\hspace{.1pt}\discretionary{.}{%
}{.}\hspace{.4pt}1637}}}


\bibitem{wright2013digital}
S.~N. Wright, P.~Kochunov, F.~Mut, M.~Bergamino, K.~M. Brown, J.~C. Mazziotta,
  A.~W. Toga, J.~R. Cebral, and G.~A. Ascoli.
\newblock Digital reconstruction and morphometric analysis of human brain
  arterial vasculature from magnetic resonance angiography.
\newblock {\em Neuroimage}, 82:170--181, 2013.

\bibitem{yang2016blockwise}
X.~Yang, L.~Shi, M.~Daianu, H.~Tong, Q.~Liu, and P.~Thompson.
\newblock Blockwise human brain network visual comparison using nodetrix
  representation.
\newblock {\em IEEE transactions on visualization and computer graphics},
  23(1):181--190, 2016.

\end{thebibliography}
\end{document}